\documentclass[conference, compsocconf]{IEEEtran}

\usepackage{epsfig,endnotes}
\usepackage{varwidth}

\usepackage{algorithm}
\usepackage{algpseudocode}

\usepackage{tikz}

\usepackage{balance}
\usepackage{nicefrac}
\usepackage[cmex10]{amsmath}
\usepackage{braket}
\usepackage{bm}
\usepackage{mathtools}
\usepackage{multirow}
\usepackage{bigdelim}
\usepackage{mathtools}
 \usepackage{amssymb}
\usepackage{amsfonts}
\let\leq\leqslant
\let\geq\geqslant

\usepackage{indentfirst}
\usepackage{booktabs}

\usepackage{enumitem}
\usepackage{boxedminipage}
\usepackage{float}
\usepackage{graphicx}
\usepackage{caption}
\usepackage{url}
\usepackage[normalem]{ulem}
\usepackage{xpatch}
\usepackage{xspace}
\usepackage{listings}
\usepackage{mathpartir}
\usepackage{makecell}
\usepackage[hidelinks]{hyperref}
\usepackage{xcolor}
\hypersetup{
    colorlinks,
    linkcolor={red!50!black},
    citecolor={blue!50!black},
    urlcolor={blue!80!black}
}
\usepackage{cleveref}
\usepackage{textcomp}
\usepackage{framed,mdframed}
\usepackage{tablefootnote}

\usepackage{verbatim}
\usepackage{fancyvrb}
\usepackage{comment}
\usepackage{mathtools}
\usepackage{scalerel}
\usepackage{amsthm}

\usepackage{stmaryrd}
\usepackage{threeparttable}
\usepackage{cite}
\usepackage{flushend}

\newlength{\saveparindent}
\newlength{\saveparskip}
\newenvironment{tiret}{%
\begin{list}{\hspace{2pt}\rule[0.5ex]{6pt}{1pt}\hfill}{\labelwidth=15pt%
\labelsep=5pt \leftmargin=20pt \topsep=3pt%
\setlength{\listparindent}{\saveparindent}%
\setlength{\parsep}{\saveparskip}%
\setlength{\itemsep}{0pt} }}{\end{list}}

\newcommand{\tool}{{\textsc{SiRnn}}\xspace}

\newtheorem{lemma}{Lemma}

\newcommand{\namedref}[2]{\hyperref[#2]{#1~\ref*{#2}}\xspace}
\newcommand{\lemmaref}[1]{\namedref{Lemma}{lem:#1}}

\newcommand{\figureref}[1]{\namedref{Figure}{fig:#1}}
\newcommand{\tableref}[1]{\namedref{Table}{tab:#1}}
\newcommand{\stepref}[1]{\namedref{Step}{step:#1}}
\newcommand{\equationref}[1]{\namedref{Equation}{eq:#1}}

\newcommand{\sectionref}[1]{\namedref{Section}{sec:#1}}
\newcommand{\appendixref}[1]{\namedref{Appendix}{app:#1}}

\newcommand{\algoref}[1]{\namedref{Algorithm}{algo:#1}}

\definecolor{forestgreen}{rgb}{0.13, 0.55, 0.13}
\definecolor{cadmiumgreen}{rgb}{0.0, 0.42, 0.24}

\definecolor{mypink}{rgb}{1,0.2,0.4}

\newcommand{\prot}{\Pi}

\newcommand{\secpar}{\lambda}

\newcommand{\zo}{\{0,1\}}

\newcommand{\share}[3]{\langle #1\rangle^{#2}_{#3}}

\newcommand{\bbZ}{\mathbb{Z}}

\newcommand{\etal}{{\em et al}.}

\newenvironment{tffbox}{
\begin{figure}[ht!]\small
}{\end{figure}}

\algdef{SE}[SUBALG]{Indent}{EndIndent}{}{\algorithmicend\ }%
\algtext*{Indent}
\algtext*{EndIndent}

\newcommand{\party}[1]{P_{#1}}
\newcommand{\func}{\mathcal{F}}
\newcommand{\msnzb}{\mathsf{MSNZB}}
\newcommand{\msnzbproj}{\mathsf{MSNZB\text{-}P}}

\newcommand{\xor}{\oplus}

\newcommand{\mill}{\mathsf{Mill}}
\newcommand{\protmill}[1]{\prot_{\mill}^{#1}}
\newcommand{\fmill}[1]{\func_{\mathsf{Mill}}^{#1}}

\newcommand{\protmilleq}[1]{\prot_{\mathsf{Mill\&Eq}}^{#1}}
\newcommand{\fand}{\func_{\mathsf{AND}}}

\newcommand{\fixed}[3]{\mathsf{Fix}\left(#1, #2, #3\right)}
\newcommand{\sigint}{\mathsf{int}}
\newcommand{\usigint}{\mathsf{uint}}
\newcommand{\flt}[3]{\mathsf{srt}_{(#2, #3)}(#1)}
\newcommand{\uflt}[3]{\mathsf{urt}_{(#2, #3)}(#1)}

\newcommand{\cA}{\mathcal{A}}
\newcommand{\cF}{\mathcal{F}}
\newcommand{\cS}{\mathcal{S}}
\newcommand{\cZ}{\mathcal{Z}}

\newcommand{\fwrap}[1]{\func_{\mathsf{Wrap}}^{#1}}
\newcommand{\feq}[1]{\func_{\mathsf{Eq}}^{#1}}
\newcommand{\fwrapeq}[1]{\func_{\mathsf{Wrap\&All1s}}^{#1}}
\newcommand{\fmux}[1]{\func_{\mathsf{MUX}}^{#1}}

\newcommand{\fBtoA}[1]{\func_{\mathsf{B2A}}^{#1}}

\newcommand{\fLUT}[1]{\func_{\mathsf{LUT}}^{#1}}

\newcommand{\bit}{\ensuremath{\{ 0, 1 \} }\xspace}
\newcommand{\kkot}[2]{\ensuremath{{#1 \choose 1}\text{-}\mathsf{OT}_{#2}}\xspace}
\newcommand{\iknpcot}[1]{\ensuremath{{2 \choose 1}\text{-}\mathsf{COT}_{#1}}\xspace}

\newcommand{\zxt}{\mathsf{ZExt}}
\newcommand{\sxt}{\mathsf{SExt}}
\newcommand{\ZX}[3]{\zxt(#1, #2, #3)}
\newcommand{\SX}[3]{\sxt(#1, #2, #3)}
\newcommand{\fzxt}[1]{\func_{\zxt}^{#1}}
\newcommand{\protzxt}[1]{\prot_{\zxt}^{#1}}
\newcommand{\fsxt}[1]{\func_{\sxt}^{#1}}
\newcommand{\protsxt}[1]{\prot_{\sxt}^{#1}}
\newcommand{\protmsnzb}[1]{\prot_{\msnzb}^{#1}}

\newcommand{\ggL}{{\gg_{\scaleto{L}{3.7pt}}}}
\newcommand{\ggA}{{\gg_{\scaleto{A}{3.7pt}}}}
\newcommand{\TR}[2]{\mathsf{TR}(#1, #2)}
\newcommand{\divtwo}[2]{\mathsf{DivPow2}(#1, #2)}

\newcommand{\flrs}[1]{\func_{\mathsf{LRS}}^{#1}}
\newcommand{\protlrs}[1]{\prot_{\mathsf{LRS}}^{#1}}
\newcommand{\fars}[1]{\func_{\mathsf{ARS}}^{#1}}
\newcommand{\protars}[1]{\prot_{\mathsf{ARS}}^{#1}}
\newcommand{\ftr}[1]{\func_{\mathsf{TR}}^{#1}}
\newcommand{\protr}[1]{\prot_{\mathsf{TR}}^{#1}}
\newcommand{\fdivtwo}[1]{\func_{\mathsf{DivPow2}}^{#1}}
\newcommand{\protdivtwo}[1]{\prot_{\mathsf{DivPow2}}^{#1}}

\newcommand{\umult}[1]{\ast_{#1}}

\newcommand{\umatmul}[1]{\boxtimes_{#1}}

\newcommand{\fumult}[1]{\func_{\mathsf{UMult}}^{#1}}
\newcommand{\fsmult}[1]{\func_{\mathsf{SMult}}^{#1}}
\newcommand{\fsmultTR}[1]{\func_{\mathsf{SMultTR}}^{#1}}

\newcommand{\fcross}[1]{\func_{\mathsf{CrossTerm}}^{#1}}
\newcommand{\fmatcross}[1]{\func_{\mathsf{MatCrossTerm}}^{#1}}
\newcommand{\fumatmul}[1]{\func_{\mathsf{UMatMul}}^{#1}}

\newcommand{\fblah}[1]{\func_{\mathsf{BitMatMul}}^{#1}}

\newcommand{\protumult}[1]{\prot_{\mathsf{UMult}}^{#1}}
\newcommand{\protsmult}[1]{\prot_{\mathsf{SMult}}^{#1}}
\newcommand{\protcross}[1]{\prot_{\mathsf{CrossTerm}}^{#1}}

\newcommand{\protumatmul}[1]{\prot_{\mathsf{UMatMul}}^{#1}}

\newcommand{\fdigdec}[1]{\func_{\mathsf{DigDec}}^{#1}}
\newcommand{\fmsnzb}[1]{\func_{\msnzb}^{#1}}
\newcommand{\fzeros}[1]{\func_{\mathsf{Zeros}}^{#1}}
\newcommand{\fonehot}[1]{\func_{\mathsf{One\text{-}Hot}}^{#1}}
\newcommand{\fmsnzbproj}[1]{\func_{\msnzbproj}^{#1}}
\newcommand{\protdigdec}[1]{\prot_{\mathsf{DigDec}}^{#1}}
\newcommand{\carry}{\mathsf{carry}}

\newcommand{\fmath}[1]{\func_f^{#1}}

\newcommand{\fexp}[1]{\func_{\mathsf{rExp}}^{#1}}

\newcommand{\bbR}{\mathbb{R}}
\newcommand{\rexp}{\mathsf{rExp}}
\newcommand{\sigmoid}{\mathsf{sigmoid}}
\newcommand{\fsigmoid}[1]{\func_\sigmoid^{#1}}
\newcommand{\fh}[1]{\func_h^{#1}}
\newcommand{\frecip}[1]{\func_{\mathsf{Rec}}^{#1}}

\newcommand{\mytanh}{\mathsf{Tanh}}
\newcommand{\ftanh}[1]{\func_{\mytanh}^{#1}}
\newcommand{\rsqrt}{\mathsf{rsqrt}}
\newcommand{\frsqrt}[1]{\func_{\rsqrt}^{#1}}

\newcommand{\sint}{m}
\newcommand{\lint}{n}
\newcommand{\digit}{d}
\newcommand{\sintx}{M}
\newcommand{\lintx}{N}

\newcommand{\alice}{\ensuremath{P_0}\xspace}
\newcommand{\bob}{\ensuremath{P_1}\xspace}
\newcommand{\ind}[1]{\ensuremath{\mathbf{1}\{ #1 \}}\xspace}
\newcommand{\concat}{||}

\newcommand{\intx}{L}

\newcommand{\wrap}{\mathsf{wrap}}

\newcommand{\naive}{\text{na\"ive}\xspace}

\newcommand{\eg}{\text{e.g.}\xspace}

\newcommand{\msb}{\mathsf{MSB}}
\newif\iffullversion

\makeatletter
\newcommand\theHALG@line{\thealgorithm.\arabic{ALG@line}}
\makeatother

\begin{document}
\title{\tool: A Math Library for Secure RNN Inference} %

\makeatletter
\newcommand{\linebreakand}{%
  \end{@IEEEauthorhalign}
  \hfill\mbox{}\par
  \mbox{}\hfill\begin{@IEEEauthorhalign}
}
\makeatother

\newcommand\blfootnote[1]{%
  \begingroup
  \renewcommand\thefootnote{}\footnote{#1}%
  \addtocounter{footnote}{-1}%
  \endgroup
}

\author{
  \IEEEauthorblockN{Deevashwer Rathee\IEEEauthorrefmark{1}}
  \IEEEauthorblockA{Microsoft Research\\
  deevashwer@berkeley.edu}
  \and
  \IEEEauthorblockN{Mayank Rathee\IEEEauthorrefmark{1}}
  \IEEEauthorblockA{Microsoft Research\\
  mayankr@berkeley.edu}
  \and
  \IEEEauthorblockN{Rahul Kranti Kiran Goli}
  \IEEEauthorblockA{Microsoft Research\\
  t-grahulk@microsoft.com}
  \linebreakand %
  \IEEEauthorblockN{Divya Gupta}
  \IEEEauthorblockA{Microsoft Research\\
  divya.gupta@microsoft.com}
  \and
  \IEEEauthorblockN{Rahul Sharma}
  \IEEEauthorblockA{Microsoft Research\\
  rahsha@microsoft.com}
  \and
  \IEEEauthorblockN{Nishanth Chandran}
  \IEEEauthorblockA{Microsoft Research\\
  nichandr@microsoft.com}
  \and
  \IEEEauthorblockN{Aseem Rastogi}
  \IEEEauthorblockA{Microsoft Research\\
  aseemr@microsoft.com}
}

\maketitle

\begin{abstract} \blfootnote{\IEEEauthorrefmark{1} Equal contribution.} \hspace*{-0.5em}
Complex machine learning (ML) inference algorithms like recurrent neural networks (RNNs) use standard functions from math libraries like exponentiation, sigmoid, tanh, and reciprocal of square root.
Although prior work on  secure 2-party inference provides specialized protocols for convolutional neural networks (CNNs), existing secure implementations of these math operators rely on generic 2-party computation (2PC) protocols that suffer from high communication. 
We provide new  specialized 2PC protocols for math functions
that crucially rely on lookup-tables and mixed-bitwidths
to address this performance overhead;
our protocols for math functions communicate up to $423\times$ less data than prior work.
Some of the mixed bitwidth operations used by our math implementations are (zero and signed) extensions, different forms of truncations, multiplication of operands of mixed-bitwidths, and digit decomposition (a generalization of bit decomposition to larger digits). For each of these primitive operations, we construct specialized 2PC protocols that are more communication efficient than generic 2PC, and can be of independent interest.
Furthermore, our math implementations are numerically precise, which ensures that the secure implementations preserve model accuracy of cleartext. 
We build on top of our novel protocols to build \tool, a library for end-to-end secure 2-party DNN inference, that provides the first secure implementations of an RNN operating on time series sensor data, an RNN operating on speech data, and a  state-of-the-art  ML architecture that combines CNNs and RNNs for identifying all heads present in  images.  
Our evaluation
shows that \tool achieves up to three orders of magnitude of performance improvement when compared
to inference of these models using an existing state-of-the-art 2PC framework.


\end{abstract}

\section{Introduction}

In the problem of secure inference, there are two parties: a server that holds a proprietary machine learning (ML) model
and a client that holds a private input. The goal is for the client to learn the prediction that the model provides on the input, with the server learning nothing about the client's input and the client learning nothing about the server's model beyond
what can be deduced from the prediction itself. Theoretically, this problem can be solved by generic secure 2-party computation (2PC)~\cite{yao,gmw}. 
Recently, this area has made great strides with the works of~\cite{cryptflow,cryptflow2,delphi,gazelle,quantizednn,nhe,nhe2,
chet,cryptonets,chameleon,cryptodl,ramparts,chimera,cingulata,
tfhe,secureml,minionn,ezpc,hycc,xonn,securenn,aby3,mpml,deepsecure,crypten,nitin}  that have made it possible to run secure inference on deep neural networks (DNNs). Frameworks for secure inference like nGraph-HE~\cite{nhe,nhe2}, MP2ML~\cite{mpml}, CrypTFlow~\cite{cryptflow,cryptflow2}, and SecureQ8~\cite{quantizednn}  go one step further and can automatically compile models trained in TensorFlow/PyTorch/ONNX to 2-party or 3-party computation  protocols secure against semi-honest adversaries.

While such systems cover the secure inference of some famous Convolutional Neural Networks (CNNs) (e.g. ResNet~\cite{resnet}, DenseNet~\cite{densenet} and MobileNet~\cite{mobilenetv2}) that exclusively use simple non-linear functions such as ReLU and Maxpool, other important architectures such as Recurrent Neural Networks (RNNs) or architectures that combine RNNs and CNNs~\cite{rnnpool} use math functions, such as exponentiation, reciprocal square root, sigmoid and tanh, extensively. 
These RNN-based architectures are the models of choice when dealing with sequential or time series data like speech~\cite{Google-30,lstm,GRU}.
Hence, for widespread adoption of secure inference, especially in the RNN application domains, a robust support for  math functions is of paramount importance.

We focus on 2-party inference secure against semi-honest adversaries\footnote{We relegate comparisons with works that need additional parties for security, e.g., 3-party computation (3PC) to~\sectionref{related}.}.
 In this setting,
works that implement math functions fall into three categories. First,
works that develop general purpose math libraries~\cite{AlyS19,mpspdz} using high-degree polynomials. %
Second, works that use boolean circuits to implement math functions~\cite{deepsecure}.
Third, works that use ad hoc piecewise linear approximations~\cite{minionn} that require developer
intervention for each dataset and each model to balance accuracy and
latency, an unacceptable ask in the context of automated frameworks
for secure inference.
All of these three approaches rely on 2PC protocols from~\cite{mpspdz,yao,aby} and suffer from huge performance overheads.

In this work, we design math functionalities that are both provably precise and efficiently realizable via novel 2PC protocols that we have developed.
The performance of all 2PC implementations depend critically on the {\em bitwidth}.
While prior works use a uniform bitwidth for the whole inference, our math functionalities use non-uniform (or mixed) bitwidths: they
operate in low bitwidths and go to high bitwidths only when necessary. 
Hence, we have developed new protocols that enable switching between bitwidths and operating on values of differing bitwidths.
Our 2PC protocols for math functionalities have upto $423\times$ lower communication than prior works (\sectionref{micro}).
We have implemented these in \tool\footnote{Read as ``siren'', \tool stands for Secure Inference for RNNs.}, a library for end-to-end DNN inference, and evaluated on RNN-based models.
While we focus on math functions occuring in RNNs, our recipe for designing math functionalities is general and can be used in other contexts. Furthermore, our math functionalities and non-uniform bitwidth protocols can also be used in non-RNN contexts and are of independent interest.

\subsection{Results in detail}

\noindent\textbf{New approximations for math functions.}
In this paper, we provide {\em provably precise} functionalities, i.e. cleartext implementations, for
exponentiation, sigmoid, tanh, and reciprocal of square root, that have been designed to minimize cryptographic overheads.
Exponentiation is used in RBF kernels~\cite{hastie}, sigmoid and tanh in RNNs with LSTM~\cite{lstm} and GRU~\cite{GRU} cells, 
and reciprocal square root in  {\tt L2Normalization}, where a vector $u$ is scaled down to a unit vector by multiplying each entry of $u$ by $\frac{1}{\sqrt{u^Tu}}$.
In a sharp departure from prior work in 2PC, our functionalities  follow the well-known paradigm of using lookup tables (LUT) to get a good initial
approximation of the math function followed by an iterative
algorithm such as Goldschmidt's iterations~\cite{goldschmidt} to improve upon
this approximation.
We take inspiration from embedded systems~\cite{ito-lookup,seedot,shiftry,wong1995fast} where the
goal of minimizing memory consumption has led to efficient low-bitwidth
implementations based on fixed-point arithmetic. Our functionalities manipulate variables with different
bitwidths to maintain precision while using minimal bitwidths. 
Furthermore, we formally verify that our functionalities provide precision guarantees similar to those provided by standard math libraries (\sectionref{validate}). \\

\noindent\textbf{Novel 2PC Protocols.} 
We provide efficient protocols for bitwidth switching (both extensions and truncations) and operating on values with differing bitwidths
so that our secure implementations mimic the behavior of the cleartext math functionalities that operate on non-uniform minimal bitwidths.
As a baseline, another option is to use existing 2PC protocols that work with a uniform bitwidth (for all values) that is large enough to accommodate all
intermediate values, i.e., avoids integer overflows. Similar to prior works, this would force us to work over much larger rings such as $\bbZ_{2^{64}}$.
Since the complexity of secure protocols grows proportionally with the bitwidth used, our use of non-uniform bitwidth leads to much more communication efficient protocols than the \naive approach of uniform bitwidth.
We consider 4 main building blocks to achieve this: (a) Extension - to increase bitwidths,
 (b) Truncation - to decrease bitwidths (and precision), (c) Multiplication - to multiply an $m$ and $n$ bit integer into an $(m+n)$-bit output to avoid
overflows (this product is later truncated to have the right bitwidth
required for further operations), and (d) Digit decomposition -
to extract relevant substrings (that we call {\em digits}) of the
input bitstring using which table lookups are performed. 
Moreover, the fixed-point cleartext code of our benchmarks also uses non-uniform bitwidths in linear layers such as matrix multiplications and convolutions, and we use our protocols for efficient realizations of the same. \\

\noindent\textbf{Secure Inference Library.} We have implemented our protocols for math functions in a new
library, called \tool\footnote{Implementation is available at \url{https://github.com/mpc-msri/EzPC}.}, for DNN
inference.
We evaluate \tool on three state-of-the-art models that use fixed-point arithmetic with non-uniform bitwidths~\cite{shiftry}.
Two of the models, one for the standard Google-30 dataset and the other for sports training,
use an RNN architecture that provides accurate
analysis of time series data~\cite{fastgrnn}. For the Google-30
dataset, the task is to
recognize commands like ``Yes'' and ``No'' from speech data,
whereas the sports training model provides 
performance feedback to a sportsperson from sensor readings.
To the best of our knowledge, this is the first empirical
evaluation of secure inference of RNNs on time series inputs like speech and
sensor readings. 
While it is possible to perform this inference using generic 2PC protocols, the overheads are intractable.
To evaluate this quantitatively, we implemented 
our benchmarks using the state-of-the-art ABY~\cite{aby} framework and this baseline is three orders of magnitude worse in latency and communication.

Our third model uses an architecture that combines RNNs and
 CNNs for the task of finding  human heads in
 images~\cite{rnnpool}. 
This model uses the reciprocal square root function that is not supported by any of the prior works on secure inference. 
Additionally, it makes roughly 3 million calls to sigmoid and tanh each.
In contrast, prior works on secure inference evaluated on models with less than 3000 calls to sigmoid/tanh~\cite{minionn,deepsecure}.
\tool can run the Heads model securely in under 7 minutes. 

\noindent To summarize, we make three key contributions:
\begin{enumerate}
\item We provide cryptographically friendly new approximations to math functions exponential, sigmoid, tanh and reciprocal square root that are provably precise (\sectionref{mathlib-protocols}).
\item We provide novel 2PC protocols for non-uniform bitwidths (\sectionref{bb}) that realize these math functionalities efficiently (up to $423\times$ lower communication than prior work, \sectionref{micro}).
\item We implement these secure implementations in the library \tool that provides the first secure inference of RNNs on speech and time series sensor data and a model that combines RNNs and CNNs. \tool outperforms state-of-the-art by three orders of magnitude in size of benchmarks (given by number of calls to math functions), latency and communication (\sectionref{eval-cases}).
Furthermore, because of the high numerical precision of our math implementations, \tool has no loss in model accuracy.
\end{enumerate}

The rest of the paper is organized as follows. We first provide a motivating example and an overview of our technical results in~\sectionref{ex}. After discussing the necessary background in~\sectionref{prelims},
we provide our novel protocols in~\sectionref{bb}. The math functionalities are discussed in~\sectionref{mathlib-protocols} with their formal verification in~\sectionref{validate}.
\sectionref{eval} provides our evaluation on microbenchmarks, i.e., math functions in isolation (\tableref{sigmoid-micro}\& \tableref{recsqrt-exp-micro}), DNNs used by prior work that use math functions (\tableref{prior-dnn}), and our RNN-based benchmarks (\tableref{rnns}).
Finally, we discuss other related work in \sectionref{related}.

\section{Overview}
\label{sec:ex}

We now present an overview of our approximations for math functions and the building block protocols required to realize them. We begin with a motivating example of an inference task that crucially uses math functions; this will help us highlight concepts such as scale and bitwidth changes.
\\\\
\noindent{\bf Motivating example.} Support vector machines (SVMs) are one of the most widely used
classical ML algorithms. While prior work on secure inference has used SVMs
 with polynomial kernels~\cite{epic,randosvm, kdd6, tdsc14} (that helps SVMs perform classification in exponentially large dimensions), the more powerful and hence  widely used 
Radial Basis Function (RBF) kernels (that operate on infinite dimensions) \cite{hastie} crucially relies on computing exponentiations, i.e., $e^x, x<0$. No prior work on secure 2PC inference supports RBF.

Consider the simple task of predicting rain using a feature
vector $x\in\mathbb{R}^2$, where $x[0]$ and $x[1]$ are temperature and
humidity respectively, and the output is yes ($y=-1$) or no ($y=1$).
An SVM with RBF model infers the result using
\[
\mathsf{sign}\left(\sum_{i=1}^k c_i e^{-\gamma^2||W_i-x||^2}\right) 
\]
where the vectors $W_i\in\mathbb{R}^2$ are part of the model and $c_i\in\{-1,1\}$.
Here, $||W_i-x||^2$ is the square of the L2 norm or the Euclidean
distance between $W_i$ and $x$. Let $k=2$, $\gamma=1$, $c_0=1$
and $c_1=-1$.
\\\\
\noindent{\bf Scales and bitwidths.} Since 2PC is much more efficient over integers than floating-point~\cite{Catrina19,cryptflow}, automated float-to-fixed converters~\cite{seedot,shiftry,fccm03,hpca99, cases02,nayak01} can be used to express this model as computation over integers 
using fixed-point arithmetic.
\begin{figure}
\begin{scriptsize}
\begin{verbatim}
int16[2][2] W = ... ; int16[2] x = ...; 
int16[2][2] U; int32[2] V; int32[2][2] T; int32[2] S; 
U[0][0] = W[0][0]-x[0];  U[0][1] = W[0][1]-x[1]; , ... 
T[0][0] = U[0][0]*U[0][0], ...
V[0] =  ((T[0][0] >> 12) + (T[0][1] >> 12), ... 
S[0] = exp(-V[0], 32, 12, 32, 30), ...
return sign(S[0] - S[1])
\end{verbatim}
\end{scriptsize}
\caption{Fixed-point code for SVM with RBF kernel}
\label{fig:fixsvm}
\end{figure}
In fixed-point arithmetic, $r \in \bbR$ is
(approximately) represented using an $\ell$-bit integer $\lfloor
r\cdot 2^s\rfloor \bmod{2^\ell}$, where $\ell$ is the {\em bitwidth} and $s\in\bbZ$ is the \emph{scale}.
Hence, fixed-point integer $a$ with scale $s$ denotes $\frac{a}{2^s} \in \bbR$.

Consider the fixed-point code for our example given  in Figure~\ref{fig:fixsvm} generated by a float-to-fixed converter.
The code stores the input $x$ and the model parameters $W$ as $16$-bit integers with scale $12$
(scale $12$ is a common setting used in several prior works on secure inference~\cite{secureml,cryptflow,cryptflow2}).
To compute the inference result,  it first computes
$U_i=W_i-x$ where $U$ has scale $12$ using standard integer
subtraction.
Next, it computes $T=U\odot U$, where $\odot$ is pointwise
multiplication. Since $U$ has $16$-bit entries, to avoid integer
overflows,
the entries of $T$ must be $32$-bits wide. Standard integer
multiplication accumulates the scale and hence entries in $T$ have a scale
of $24$.
Thus, the code right shifts the entries of $T$ by $12$ to bring the scale back
to $12$ and accumulate them in $V_i=||W_i-x||^2$.
Next, it calls exponentiation on {\em negative} inputs of bitwidth $32$ and
scale $12$ and produces the result $S$ with bitwidth $32$ and scale $30$.
The final result is the sign of $c_0S[0]+c_1S[1]$. 
\tool incurs less than 30KB of communication to run this code.

Observe that the fixed-point code in \figureref{fixsvm} frequently changes bitwidths and scales with each operation.
As we describe in \figureref{exp} (\sectionref{mathlib-protocols}), our math functionality for exponential would require multiplying two $32$-bit values to compute an intermediate $64$-bit result.
Now, if we had to implement \figureref{fixsvm} using existing 2PC protocols, we would be forced to use uniform bitwidth of at least $64$ for all variables.
In particular, the bitwidths of $x, W, U, T, V, S$ will all be $64$ instead of $16$ or $32$. 
More generally, the requirement of a high bitwidth even in one operation, coupled with the requirement of uniform bitwidths, raises the bitwidths of all variables and operations throughout an inference task, resulting in a communication blowup.
In contrast \tool provides novel protocols for these low-level operations of switching bitwidth and scale and multiplying values of small bitwidth into large bitwidth. Ensuring that bitwidths used in secure code mimic the bitwidths used in low-bitwidth cleartext code, is the key factor
in \mbox{low communication complexity of our secure math functions}.

Next we give an overview of our approximations for math functions followed by building blocks for our protocols.

\subsection{Our approximations for math functions}
Our math functionalities are designed keeping cryptographic costs in mind. We first use lookup tables (LUT) to get a
good initial approximation of the math functions and then run an iterative algorithm such as Goldschmidt's iterations to improve upon this approximation. 
Larger LUTs lead to more precise results.
However, the communication of secure protocol for LUTs grows linearly with size of LUT.
Hence, we need to strike a balance to obtain implementations that are both precise and communication efficient. 
Thus, for exponentiation for negative inputs, we break the input bitstring $x$ into smaller $d$-length substrings (via digit decomposition) that are used to index
multiple $2^d$-sized LUTs. The looked up values are multiplied into high bit intermediate results which are then truncated to match the specified output bitwidth and scale.

Sigmoid and tanh reduce to exponentiating negative values and reciprocating values between $1$ and $2$. 
For the latter, Ito \etal \cite{ito-lookup} provide a method for initial
approximation of reciprocal using an LUT.  After obtaining an initial approximation with
$\ell$ bit entries and $\ell-2$ bits of fractional part, we iterate
using standard Goldschmidt's method. To make these iterations
communication efficient, we run them using fixed-point arithmetic with non-uniform
bit-widths. Our implementation for reciprocal square root is similar but requires
additional work to shift the initial input to be between $1$ and $2$
using the most significant non-zero bit (MSNZB).

\subsection{2PC protocols in \tool}

The 2PC protocols in \tool are based on 4 building blocks: (a) Extension; (b) Truncation; (c) Multiplication; 
and (d) Digit decomposition.
Our protocols mimic the low bitwidths used by cleartext fixed-point code, and work over power-of-2 rings, i.e. $\bbZ_{2^\ell}$. %
Let 
$\secpar = 128$ be the computational security parameter.

\paragraph{Extension}
This is used to lift values
from smaller ring $\bbZ_{2^m}$ to larger ring $\bbZ_{2^n}$ (i.e. $m <
n$).
Although extension has been considered in honest majority three-party computation~\cite{sharemind-randmets}, there are no specialized 2PC protocols for it.
A natural baseline, however, is provided by Yao's garbled circuits\footnote{Depth optimized GMW~\cite{gmw} has higher communication than GC for our functionalities.} (GC) \cite{yao}, which requires around $\secpar(4m+2n)$ bits of communication to
reconstruct and re-share.
In contrast, our protocol requires around
$\secpar m$ bits of communication, that is roughly $6\times$ better than GC.

\paragraph{Truncation} This operation is used to reduce scale and is often used after multiplication.
 We require 4 kinds of truncation operations for $\ell$-bit values by $s$ bits:
logical and arithmetic right shifts (that preserve the
bitwidth), truncate-and-reduce  (outputs the truncated value in
$\bbZ_{2^{\ell-s}}$), and division by $2^s$.
State-of-the-art protocol
for arithmetic right shift (ARS) was given by \cite{cryptflow2} with
communication roughly $\secpar(\ell+s)$ that can also be used for logical right shift and truncate-and-reduce. 
We give a new protocol for logical/arithmetic right shift with communication $\approx \secpar \ell$, i.e., independent of $\secpar s$.
Moreover, most of our math functionalities require only truncate-and-reduce that decreases both scale and bitwidth.
We show how to achieve this in only $\approx \secpar(s+1)$ bits of communication. 
Finally,
our fixed-point benchmarks also require a division by power-of-2
operation that is different from ARS for negative $x$ and outputs
$\lceil x/2^s \rceil$. Our protocol for this division requires roughly
$4.5\times$ less communication than GC.

\paragraph{Multiplication} We consider the functionality  for multiplying an $m$-bit integer with an $n$-bit integer to produce an $\ell = (m+n)$-bit output. This choice of $\ell$ ensures that there are no overflows.
A  similar functionality  has been considered in the
3-party setting~\cite{sharemind-randmets} that extends both operands
to $\ell$ bits and then invokes existing multiplication protocols over
$\ell$ bits. This approach can be used in 2PC setting as well using
our optimized protocols for extension (that are $6\times$ better than GC). We provide
an alternate protocol that requires $1.5\times$ less communication than
the \naive approach of extend-then-multiply.

\paragraph{Digit Decomposition} This splits an $\ell$-bit value into $c = \ell/d$ digits of $d$-bits.
It can be realized using GC with communication $\secpar (6\ell - 2c - 2)$ bits. 
We propose an optimized protocol that 
requires communication of $\approx \secpar(c-1)(d+2)$ bits, that is, roughly $5\times$ lower than GC.
We build on digit decomposition for an efficient protocol for MSNZB required to realize the functionality for reciprocal square root.

\section{Preliminaries}
\label{sec:prelims}

\subsection{Math functions and ULP errors} 
 The math functions have irrational outputs which are impossible to represent exactly in finite number of bits. 
When using a finite-bit representation, like
floating-point or fixed-point, the most precise implementation is the
one that generates {\em correctly rounded} results, i.e., the output
of the implementation is a representable number that is closest to
the ideal $\mathbb{R}$ result.
However, because of Table maker's dilemma, such implementations are
computationally very expensive~\cite{mpfr}.
Consequently, standard math libraries like GNU's or Intel's {\tt libm}
don't return the correctly rounded results.

{\em ULP error.}  
The deviation between the finite-bit output and the exact result can be quantified in three ways: absolute error, relative error, and ``units in last place'' or ULPs. The former two have serious issues and the ``most natural way to measure rounding error is in ulps''~\cite{goldberg}; standard math libraries use ULPs to report the precision of their implementations~\cite{mkl,svml}. To see why this is the case, observe that if $r$ is a very small real number, then the absolute error between $r$ and $r'=2r$, i.e., $|r-r'|=|r|$, is small as well. Hence, a low absolute error can be achieved even when every bit of the output is incorrect. Relative error, given by $|\frac{r-r'}{r}|$, remedies this situation and $r'=2r$ leads to high relative errors irrespective of the magnitude of $r$. However, the relative error is undefined for $r=0$. ULP errors have the nice property that they are always well-defined and don't grow or shrink with the magnitude of $r$. At a high level, the ULP error\footnote{See~\cite{goldberg} for the formal definition of ULPs.} between an exact real result $r$ and the library output $a$ is the number of representable numbers between $a$ and $r$~\cite{pldi16,pldi14}. We show an example in Figure~\ref{fig:ulp}. %

Intel's SVML~\cite{svml} has ULP
error below $4$ and MKL~\cite{mkl}  guarantees ULP error below $1$.
It is important for the ULP error to be low for reusability of the
library implementations as a low error 
gives the developers an assurance that
the library is
producing precise results inasmuch as the underlying
representation permits.

\begin{figure}
\centering
  \includegraphics[scale=0.4]{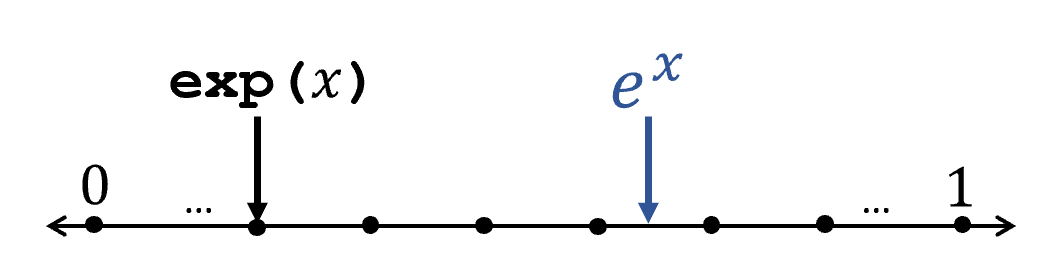}
  \caption{The computed result $\mathtt{exp(x)}$ is in error of 3 ULPs from the mathematically exact result $e^x$. Dots denote the representable numbers.}
  \label{fig:ulp}
\end{figure}

\subsection{Threat Model}
We consider 2-party computation secure against a {\em static semi-honest} adversary running in probabilistic polynomial time.
That is, we consider a computationally bounded adversary $\mathcal{A}$ that corrupts one of the parties at the beginning of the protocol execution, follows the protocol specification, but tries to learn additional information about the honest party's input. We argue security using the simulation paradigm~\cite{gmw, canetti00, lindellsim}.
For any function $f$ to be computed, consider following two interactions: a real interaction where $P_0$ and $P_1$ interact using the protocol specification in the presence of $\cA$ and the environment $\cZ$
and the ideal interaction where $P_0, P_1$ send their inputs to the trusted functionality $\cF$ that computes $f$ and sends the outputs to the parties. 
We argue that for every real adversary $\cA$, there is an ideal adversary $\cS$ such that no environment $\cZ$ interacting externally with the adversaries can distinguish between real and ideal interactions. Our protocols invoke several sub-protocols and for ease of exposition we describe them using the {\em hybrid model}, which is the same as a real interaction except that the sub-protocol executions are replaced with calls to the corresponding trusted functionalities -- protocol invoking $\cF$ is said to be in the $\cF$-hybrid model. %

\subsection{Notation}
\label{sec:notation}
Let $\secpar$ be computational security parameter.
Uppercase $L, M, N$ denote $2^\ell, 2^\sint, 2^\lint$, respectively.
$[k]$ refers to the set $\{0,\ldots,k-1\}$. 
$\ind{b}$ denotes the indicator function that is 1 when $b$ is true, and 0 otherwise. 
We use the natural one-to-one correspondence between $\zo^\ell$ and $\bbZ_L$.
Consider the lossless lifting operators $\zeta_\ell$ that maps an element of ring $\bbZ_L$ to $\bbZ$  and $\zeta_{\ell,m}$ for $m\geq\ell$ that maps an element of ring $\bbZ_L$ to $\bbZ_M$. 
For brevity, we suppress these operations when their unambiguous use can be deduced from the context.
For an element $x \in \bbZ_{\intx}$, $\sigint(x)$ and $\usigint(x)$ refer to the signed and unsigned values in $\bbZ$ respectively, where the signed case corresponds to the 2's complement representation. $\usigint(x)$ is defined as $\zeta_\ell(x)$ and  $\sigint(x) = \usigint(x) - \msb(x) \cdot \intx$, where $\msb(x) = \ind{x \geq 2^{\ell - 1}}$ is the most significant bit. 
For $x, y \in \bbZ_L$, $\wrap(x, y, L)$ is 1 if $x + y \geq L$ over $\bbZ$ and 0 otherwise. 
Finally, consider the operator $\umult{m}: \bbZ \times \bbZ \rightarrow \bbZ_M$ where $x \umult{m} y = x \cdot y \bmod M$.
When one or both inputs are from some integer ring $\bbZ_L$, we use $\usigint()$ and $\sigint()$ to map the element to $\bbZ$.

{\em Fixed-Point Representation.} We encode real numbers as elements in $\bbZ_{\intx}$ using their fixed-point representation. 
Fixed-point representation in $\bbZ_L$ defines 2 variables, $\ell$ and $s$, where $\ell$ is the {\em bitwidth}, $s$ is the resolution (or, fractional part bitwidth) referred to as the {\em scale} and $\ell-s$ is the bitwidth for the integer part. 
A real number $x \in \mathbb{R}$ is encoded into its fixed-point representation $\hat{x} \in \bbZ_L$ with bitwidth $\ell$ and scale $s$ as $\hat{x} = \fixed{x}{\ell}{s} =  \lfloor x \cdot 2^s \rfloor \bmod L$.
The reverse mappings from fixed-point representation to reals are 
$\uflt{a}{\ell}{s} = \usigint(a) / 2^s$ for unsigned numbers and $\flt{a}{\ell}{s} = \sigint(a) / 2^s$ for signed numbers, where division is over $\mathbb{R}$.

\subsection{Cryptographic Primitives}
\label{sec:crypto-bb}
\noindent{\em Secret Sharing.} We use 2-out-of-2 additive secret sharing schemes over different power-of-2 rings \cite{blakely79, shamir79}. For $x \in \bbZ_L$, we denote its shares by $\share{x}{\ell}{} = (\share{x}{\ell}{0}, \share{x}{\ell}{1})$ such that $x = \share{x}{\ell}{0} + \share{x}{\ell}{1} \bmod L$ and $P_b$ holds $\share{x}{\ell}{b}$ for $b \in \zo$. When $\ell = 1$, i.e., over $\bbZ_2$, we use $\share{x}{B}{}$ to denote boolean shares.
In our protocols, we write ``$P_0$ \& $P_1$ hold $\share{x}{\ell}{}$.'' to denote that $P_b$ holds $\share{x}{\ell}{b}$ for $b \in \zo$.

\noindent{\em Oblivious Transfer.}
Consider 2-party functionality 1-out-of-$k$ oblivious transfer (OT) denoted by $\kkot{k}{\ell}$, where one party is the sender with $k$ $\ell$-bit messages $x_0, \ldots, x_{k-1} \in \zo^\ell$ and the other party is the receiver with an  index $j \in [k]$. The receiver learns $x_j$ as the output, and the sender learns nothing.
We realize this functionality using the OT extension protocol from \cite{kkot}, which optimizes and generalizes the protocol from \cite{iknp}.
Additionally, we use the 1-out-of-$2$ correlated OT (COT) functionality $\iknpcot{\ell}$, which is defined as follows:
sender inputs a correlation $x~\in~\bbZ_{\intx}$, receiver inputs a choice bit $j \in \zo$, and the functionality outputs a random element $r \in \bbZ_{\intx}$ to the sender and $-r + j \cdot x$ to the receiver.
We instantiate this functionality with the COT protocol from \cite{ALSZ13}.
Excluding the one-time setup cost for the base OTs, $\kkot{k}{\ell}$ and $\iknpcot{\ell}$ require $2\lambda + k\ell$ and $\lambda + \ell$~bits of communication, respectively, and execute in $2$ rounds\footnote{Recently, MOTION~\cite{motion} gave a COT protocol with similar communication and overall 2 rounds. However, their protocol requires only a single round of communication assuming precomputed ROT correlations. The total round complexity of some of our protocols can benefit from this COT.}.
For the special case of $k=2$,  $\kkot{2}{\ell}$ requires $\lambda+2\ell$ bits of communication~\cite{ALSZ13}.

\subsection{2PC Functionalities} \label{sec:two-pc-prelims}
For a 2-party functionality $\mathcal{F}$, we say that ``$P_0$ \& $P_1$ invoke $\mathcal{F}(x, y)$ to learn $\share{z}{\ell}{}$'' to mean that $P_0$ with input $x$ and $P_1$ with input $y$ invoke $\mathcal{F}$ and learn arithmetic shares of $z$ over $\bbZ_L$, i.e., $P_0$ gets $\share{z}{\ell}{0}$ and $P_1$ gets $\share{z}{\ell}{1}$. 
We write ``$\mathcal{F}(\share{x}{\ell}{})$'' to mean that $\mathcal{F}$ takes $\share{x}{\ell}{0}$ from $P_0$ and $\share{x}{\ell}{1}$ from $P_1$.
In our protocols, we use the following 2-party functionalities.

\begin{tiret}
\item {\em Millionaires'/Wrap}: The $\ell$-bit Millionaires' functionality, $\fmill{\ell}$ takes as input $x \in \zo^\ell$ from $P_0$ and $y\in\zo^\ell$ from $P_1$ and returns $\share{z}{B}{}$ such that $z = \ind{x < y}$. 
The $\ell$-bit wrap functionality, $\fwrap{\ell}$ on same inputs returns $\share{z}{B}{}$ such that $z = \wrap(x, y, L)$. Note that $\fwrap{\ell}(x, y) = \fmill{\ell}(L-1-x, y)$. 
Recently, \cite{cryptflow2} gave an efficient protocol for $\fmill{\ell}$ with communication less than\footnote{For ease of exposition, we use this rough upper bound to compute an upper bound of communication of most of our protocols.} $\secpar\ell + 14\ell$ bits with $\log \ell$ rounds.

\item {\sf AND}: 
The functionality $\fand$ takes as input $(\share{x}{B}{}, \share{y}{B}{})$ and returns $\share{x\wedge y}{B}{}$.
$\fand$ can be realized using Beaver bit-triples~\cite{beaver} and \cite{cryptflow2} gave a protocol for $\fand$ with $\secpar+20$ or $148$ bits\footnote{The best known communication for $\fand$ is 138 bits~\cite{DKSSZZ}, however, its implementation isn't available.} of total communication.

\item {\em Boolean to Arithmetic (B2A)}: The $\ell$-bit B2A functionality, $\fBtoA{\ell}$, takes  boolean shares $\share{x}{B}{}$ and outputs arithmetic shares of the same value, i.e., $\share{x}{\ell}{}$.
We use the COT based protocol from \cite{cryptflow2} with communication $\secpar + \ell$ bits.
\item {\em Multiplexer (MUX)}: The $\ell$-bit MUX functionality, 
$\fmux{\ell}$, takes as input $\share{x}{B}{}$ and $\share{y}{\ell}{}$ and outputs $\share{z}{\ell}{}$ such that $z = y$ if $x = 1$ and 0 otherwise.
We provide an optimized protocol that reduces communication from $2(\secpar + 2\ell)$ \cite{cryptflow2} to $2(\secpar+\ell)$ (see \appendixref{mux}).

\item {\em Lookup Table (LUT)}: The  LUT functionality for table $T$ with $M$ entries of $n$-bits each, $\fLUT{T, m, n}$ takes as input $\share{x}{m}{}$ and outputs $\share{z}{n}{}$ such that $z = T[x]$.  
It can be realized using a single call to $\kkot{M}{n}$ with communication $2\secpar+Mn$ bits~\cite{DKSSZZ}.

\end{tiret}

\section{Building Block Protocols}
\label{sec:bb}
In this section, we describe our building block protocols that we combine later to obtain protocols for math library functions in \sectionref{mathlib-protocols}.
Our protocols extensively use the existing 2PC functionalities described in \sectionref{two-pc-prelims}.
In addition, they invoke the functionality $\fwrapeq{\ell}$ that takes as input $x \in \zo^\ell$ from $P_0$ and $y \in \zo^\ell$ from $P_1$ and outputs $(\share{w}{B}{} \concat \share{e}{B}{})$ such that $w = \wrap(x, y, L)$ and $e = \ind{(x + y \bmod L) = L-1}$. We show that this functionality can be realized with nearly the same cost as $\fwrap{\ell}$ by making a white-box use of the protocol for $\fmill{\ell}$ from \cite{cryptflow2} (\appendixref{wrap-eq}).
The resulting protocol has $\log\ell$ rounds and  at most $\secpar \ell + 14 \ell$ bits of communication.
Below we describe our protocols for extension, truncation, multiplication, digit decomposition and MSB-to-wrap optimization that applies extensively to our math functionalities.

\subsection{Zero Extension and Signed Extension} 
\label{sec:extension}

Zero and signed extension functions are used to extend the bitwidths of unsigned and signed numbers, respectively. More precisely, for an $\sint$-bit number $x \in \bbZ_{\sintx}$, we define zero extension (resp. signed extension) to $\lint$-bits ($\lint > \sint$) by $y = \ZX{x}{\sint}{\lint} \in \bbZ_{\lintx}$ (resp. $y = \SX{x}{\sint}{\lint} \in \bbZ_{\lintx}$), such that $\usigint(y) = \usigint(x)$ (resp. $\sigint(y) = \sigint(x)$) holds. 
In \algoref{zxt}, we describe our protocol for $\fzxt{m,n}$ that takes as input $\share{x}{m}{}$ and outputs $\share{y}{n}{}$, where $y = \ZX{x}{\sint}{\lint}$. 
This protocol requires $\log m + 2$ rounds and less than $\secpar (m + 1) + 13 m + n$ bits of communication.

Correctness of our protocol can be argued as follows: 
    By correctness of $\fwrap{m}$ and $\fBtoA{n-m}$, it holds that $w = \wrap(\share{x}{\sint}{0}, \share{x}{\sint}{1}, M)$ and  $y = \sum_{b=0}^{1} (\share{x}{m}{b} - M \cdot \share{w}{n-m}{b}) \bmod N$. 
Over $\bbZ$, $w = \share{w}{n-m}{0} + \share{w}{n-m}{1} - 2^{n-m} \cdot \wrap(\share{w}{n-m}{0}, \share{w}{n-m}{1}, 2^{n-m})$.
Thus, $M \umult{n} w = M \umult{n} (\share{w}{n-m}{0} + \share{w}{n-m}{1}) $.
Also, over $\bbZ$, $x = \share{x}{\sint}{0} + \share{x}{\sint}{1} - w\cdot M$. Hence, $x \bmod N = y$.

Our protocol for signed extension, i.e., $\fsxt{m,n}$, uses the following equation over $\bbZ$:
\begin{equation} \label{eq:sext}
    \sigint(x) = x' - 2^{m-1} \text{, for $x' = x + 2^{m-1} \bmod{M}$}.
\end{equation}
This gives\footnote{A similar relation was used in \cite{edabits} for truncation.} ${\SX{x}{\sint}{\lint} = \ZX{x'}{\sint}{\lint} - 2^{\sint-1}}$. 

As a baseline, one can use garbled circuits (GC) to realize zero and signed-extensions with communication cost of $\secpar (4 m + 2n - 4)$ bits, i.e., roughly $6\times$ the cost of our protocols.

\begin{algorithm}[t]
\caption{Zero Extension, $\protzxt{\sint,\lint}$:}
\label{algo:zxt}
\begin{algorithmic}[1]
\small

    \Require $P_0$ $\&$ $P_1$ hold $\share{x}{\sint}{}$.
    \Ensure $P_0$ $\&$ $P_1$ get $\share{y}{\lint}{}$ for $y = \ZX{x}{\sint}{\lint}$.

\vspace{0.1cm}

    \State \alice \& \bob invoke $\fwrap{\sint}(\share{x}{\sint}{0}, \share{x}{\sint}{1})$ and learn $\share{w}{B}{}$.
    \State  \alice \& \bob invoke $\fBtoA{\lint-\sint}(\share{w}{B}{})$ and learn $\share{w}{\lint-\sint}{}$.
\State For $b \in \zo$, $\party{b}$ outputs $\share{y}{\lint}{b} = \share{x}{\sint}{b} - \sintx \umult{\lint} \share{w}{\lint-\sint}{b}$.

\end{algorithmic}
\end{algorithm}

\subsection{Truncation} 
\label{sec:truncation}
We consider four types of truncation operations for ring $\bbZ_L$ as follows:
We denote the logical and arithmetic right-shift operators by $\ggL$ and $\ggA$, respectively, whose inputs are outputs are in $\bbZ_L$.
Next, we define $\TR{x}{s}$ (Truncate \& Reduce $x$ by $s$-bits) that takes inputs in $\bbZ_L$, drops  the lower $s$-bits from the bit-representation of $x$ and outputs the truncated value in smaller ring, $\bbZ_{2^{\ell-s}}$.
Additionally, our benchmarks also require the C-style division (quotients are rounded towards $0$) where the divisor is a power-of-2. 

\begin{algorithm}[h]
\caption{Logical Right Shift, $\protlrs{\ell,s}$:}
\label{algo:lrs}
\begin{algorithmic}[1]
\small

\Require $P_0$ \& $P_1$ hold $\share{x}{\ell}{}$.
\Ensure $P_0$ \& $P_1$ get $\share{x \ggL s}{\ell}{}$.

\vspace{0.1cm}

\State For $b \in \zo$, $\party{b}$ parses $\share{x}{\ell}{b}$ as an $\ell$-bit string $u_b\concat v_b$, where $u_b \in \bit^{\ell-s}$ and $v_b \in \bit^{s}$.
\State \alice \& \bob invoke $\fwrap{s}(v_0, v_1)$ and learn $\share{c}{B}{}$.
\State \alice \& \bob invoke $\fwrapeq{\ell-s}(u_0, u_1)$ and learn $\share{d}{B}{}\concat \share{e}{B}{}$. \label{step:first}
\State \alice \& \bob invoke $\fand(\share{c}{B}{},\share{e}{B}{})$ and learn~$\share{t}{B}{}$.
\State For $b \in \zo$, $\party{b}$ sets $\share{w}{B}{b} = \share{d}{B}{b} \oplus \share{t}{B}{b}$.
\State \alice \& \bob invoke $\fBtoA{\ell}(\share{c}{B}{})$ and learn $\share{c}{\ell}{}$.
\State \alice \& \bob invoke $\fBtoA{s}(\share{w}{B}{})$ and learn $\share{w}{s}{}$. \label{step:last}
\State For $b \in \zo$, $\party{b}$ outputs $u_b - 2^{\ell-s} \umult{\ell} \share{w}{s}{b} + \share{c}{\ell}{b}$.

\end{algorithmic}
\end{algorithm}

\noindent{\em Logical Right Shift.} In \algoref{lrs}, we describe our protocol for $\flrs{\ell, s}$ that takes as input $\share{x}{\ell}{}$ and outputs $\share{x \ggL s}{\ell}{}$. The idea is as follows: Consider $x \in \bbZ_L$ and $\share{x}{\ell}{}$. Also, for $b \in \zo$, let $\share{x}{\ell}{b} = u_b \concat v_b$ where $u_b \in \zo^{\ell-s}$ and $v_b \in \zo^s$. Then, it can be shown that  $x \ggL s = u_0 + u_1 - 2^{\ell-s}\cdot \wrap(\share{x}{\ell}{0}, \share{x}{\ell}{1}, L)+ \wrap(v_0, v_1, 2^s)$ \cite{fss20}. 
A simple protocol for $\flrs{\ell,s}$  computes shares of wrap terms over $\ell$-bits and $s$-bits separately. We further optimize this protocol using the following lemma (proof appears in~\appendixref{truncation}):
\begin{lemma}
\label{lem:lrs}
    Let $x \in \bbZ_L$, $\share{x}{\ell}{}$ be shares of $x$ and for $b \in \zo$, $\share{x}{\ell}{b} = u_b \concat v_b$, where $u_b \in \zo^{\ell-s}$ and $v_b \in \zo^s$. Define $c = \wrap(v_0,v_1,2^s)$, $d = \wrap(u_0,u_1,2^{\ell-s})$, $e = \ind{u_0 + u_1 \bmod 2^{\ell-s} = 2^{\ell-s}-1}$ and $w = \wrap(\share{x}{\ell}{0}, \share{x}{\ell}{1}, L)$, then it holds that $w = d \oplus (c \land e)$. 
\end{lemma}

Using this lemma, our protocol only uses wrap computations over $\ell-s$ and $s$ bits and a call to $\fand$ functionality. As another optimization, while invoking $\fBtoA{}$ on shares of $w$, we go to arithmetic shares over $\bbZ_{2^s}$ (and not $\bbZ_L$). Overall communication cost is less than $\secpar(\ell + 3) + 15\ell + s + 20$ and rounds required are $\log\ell + 3$. \\

\noindent{\em Arithmetic Right Shift.} Our protocol for $\fars{\ell,s}$ that outputs $\share{x \ggA s}{\ell}{}$ builds upon $\flrs{\ell,s}$ using the relation\cite{edabits}: $x \ggA s = x' \ggL s - 2^{\ell-s-1}$, where $x' = x + 2^{\ell-1}$. Hence, it has the same cost as $\protlrs{\ell,s}$.
Prior state-of-the-art protocol for arithmetic right shift is from CrypTFlow2~\cite{cryptflow2} that runs in $\log\ell + 2$ rounds with communication $\secpar (\ell + s + 2) + 19\ell + 14s$ bits. 
Note that unlike our protocol, its communication grows multiplicatively in $\secpar$ with both $\ell$ and $s$. \\

\noindent{\em Truncate and Reduce.} Many of our protocols can benefit from truncate and reduce to the smaller ring over logical/arithmetic right shift operations that output shares in the original ring. At a high level, our protocol for $\ftr{\ell,s}$ that outputs $\share{\TR{x}{s}}{\ell-s}{}$ is as follows: Using the above notation, $\TR{x}{ s} = u_0 + u_1 + \wrap(v_0, v_1, 2^s)$. Hence, we can skip the computation of shares of $w$, i.e., steps~\ref{step:first}--\ref{step:last} can be skipped. Overall communication is $\secpar (s + 1) + \ell + 13s$ bits.
The best solution using prior techniques is: $\TR{x}{s} = (x \ggA s) \bmod{2^{\ell-s}}$, which would incur the same cost as the state-of-the-art ARS protocol~\cite{cryptflow2}, i.e., $\secpar (\ell + s + 2) + 19\ell + 14s$ bits.
\\

\noindent{\em Division by power-of-2.}
 In addition to arithmetic right shift, the fixed-point code for ML benchmarks require C-style division by power-of-2 to preserve model accuracy.
Consider the functionality $\fdivtwo{\ell,s}$ that takes $\share{x}{\ell}{}$ as input and outputs $\share{z}{\ell}{}$ such that $z = \lceil \sigint(x)/2^s \rceil \bmod\ L$ for $z <0$ and $z = \lfloor \sigint(x)/2^s \rfloor \bmod\ L$ for $z \geq 0$.
We give an overview of our protocol in \appendixref{truncation} that requires roughly $\secpar(\ell + 2s + 4)$ bits of communication.
To the best of our knowledge, no prior work explicitly builds a protocol for this functionality. A garbled circuits implementation, costs $\secpar (8\ell + 2s - 6)$ bits.

\subsection{Multiplication with non-uniform bitwidths} 
\label{sec:multiplication}

Our machine learning models as well as math library functions (see \sectionref{mathlib-protocols}) use  multiplication operation with operands of different bit-widths that outputs a value in the larger ring. Below, we describe these functions and their protocols for both the unsigned and the signed case. \\

\noindent{\em Unsigned Multiplication with non-uniform bitwidths.} 
Consider the functionality $\fumult{m,n}$ that takes $\share{x}{\sint}{}$ and $\share{y}{\lint}{}$ as input and returns $\share{z}{\ell}{}$, where $z = x \umult{\ell} y$, for $\ell = m+n$.
In contrast, all prior works on secure inference~\cite{secureml,minionn,gazelle,delphi,cryptflow2,deepsecure}, use $m = n = \ell$. 
A \naive\ way to realize this functionality is to first extend both the inputs to $\ell$-bits and then use standard multiplication, i.e., multiply $\fzxt{m,\ell}(\share{x}{m}{})$ and $\fzxt{n,\ell}(\share{y}{n}{})$ using existing protocols for uniform bit-widths. 
We give a new custom protocol for multiplying values of non-uniform bitwidths that beats this \naive approach by roughly $1.5 \times$. %
Our protocol builds on the functionality $\fcross{m,n} : \bbZ_M \times \bbZ_N \rightarrow \bbZ_L \times \bbZ_L$  that is defined as $\fcross{m,n}(x, y) = \share{z}{\ell}{}$, where $z = x \umult{\ell} y$. 
We describe our protocol for $\fcross{m,n}$ in \appendixref{cross} that carefully uses $\iknpcot{}$ (to minimize overall communication) similar to the techniques of generating Beaver triples~\cite{beaver}. 
The communication complexity of this protocol is $\mu (\secpar + \mu/2 + 1/2) + mn$, where $\mu = \mathsf{min}(m,n)$.

\begin{algorithm}[t]
\caption{Unsigned Multiplication, $\protumult{\sint,\lint}$:}
\label{algo:umult}
\begin{algorithmic}[1]
\small

\Require $P_0$ \& $P_1$ hold $\share{x}{\sint}{}$ and $\share{y}{\lint}{}$.

\Ensure $P_0$ \& $P_1$ get $\share{z}{\ell}{}$, where $z = x \umult{\ell} y$ and $\ell = m+n$.

\vspace{0.1cm}
\State For $b \in \zo$, let $x_b = \share{x}{\sint}{b}$ and $y_b = \share{y}{\lint}{b}$.
\State $P_0$ and $P_1$ invoke the following functionalities.
\Indent
\State $\fcross{m, n}(x_0, y_{1})$ and learn $\share{c}{\ell}{}$. \label{umult-cross-one}
\State $\fcross{n, m}(y_{0}, x_1)$ and learn $\share{d}{\ell}{}$. \label{umult-cross-two}
\State $\fwrap{m}(x_0, x_1)$ to learn $\share{w_x}{B}{}$. \label{umult-wrap-one}
\State $\fwrap{\lint}(y_0, y_1)$ to learn $\share{w_y}{B}{}$. \label{umult-wrap-two}
\State $\fmux{\sint}(\share{w_y}{B}{}, \share{x}{m}{})$ to learn $\share{g}{\sint}{}$. \label{umult-mux-one}
\State $\fmux{\lint}(\share{w_x}{B}{}, \share{y}{n}{})$ to learn $\share{h}{\lint}{}$. \label{umult-mux-two}
\EndIndent
\State $P_b$ outputs  $x_b \umult{\ell} y_b + \share{c}{\ell}{b} + \share{d}{\ell}{b} - \lintx \umult{\ell} \share{g}{\sint}{b} - \sintx \umult{\ell} \share{h}{\lint}{b}$ for $b \in \zo$.

\end{algorithmic}
\end{algorithm}

By definition of $\umult{\ell}$, we wish to compute $\usigint(x) \cdot \usigint(y) \bmod L$, where $\ell = m+n$. 
Let $x = x_0 + x_1 \bmod\ M$ and $y = y_0 + y_1 \bmod\ N$. 
\algoref{umult} gives our protocol for $\fumult{m,n}$ that builds on the following: 
Over $\bbZ$, 
\begin{align}
  \usigint(x) \cdot \usigint(y) =& ~(x_0 + x_1 - 2^{\sint} w_x) \cdot (y_0 + y_1 - 2^{\lint} w_y) \nonumber \\
  =&  ~ x_0 y_0 + x_1 y_1 + x_0 y_1 + x_1 y_0 \nonumber \\
  & ~ - 2^{\sint} w_x y - 2^{\lint} w_y x - 2^{\ell} w_x w_y, \label{eq:umult}
\end{align}
where  $w_x~=\wrap(x_0, x_1, M)$ and $w_y = \wrap(y_0, y_1, N)$. 
Taking a $\bmod\ L$, removes the last term.
In the protocol, party $P_b$ computes $x_by_b$ as $(x_b\; \umult{\ell}\; y_b)$ locally and invokes $\fcross{m,n}$ to compute shares of cross-terms $x_b  y_{1-b}$. Wraps are computed using $\fwrap{}$ and multiplied to values using $\fmux{}$.

The communication complexity of our protocol is roughly 
$\secpar(3\mu + \nu) + \mu(\mu + 2\nu) + 16(m+n)$ where $\mu = \mathsf{min}(m,n)$ and $\nu = \mathsf{max}(m,n)$. 
In contrast, communication complexity of \naive approach of extend-then-multiply that uses our optimized protocols for extension is roughly $3\secpar(\mu+\nu) + (m+n)^2 + 15(m+n)$, i.e., roughly $1.5\times$ more than our new protocol.

We note that the same ideas also work for the setting $\ell < m+n$ by using an appropriate protocol for $\fcross{m,n,\ell}$ with specific value of $\ell$.
Similarly, we define the multiplication functionality $\fumult{m,n,\ell}$ which internally invokes $\fcross{m,n,\ell}$, where the additional superscript denotes the bitwidth of the output.
Our protocols for math library functions also uses this setting for better efficiency.\\

\noindent{\em Signed Multiplication with non-uniform bitwidths.}
Consider the functionality $\fsmult{m,n}$ that takes $\share{x}{\sint}{}$ and $\share{y}{\lint}{}$ as input and returns $\share{z}{\ell}{}$, where $z = \sigint(x) \umult{\ell} \sigint(y)$, for $\ell = m+n$.
Let $x' = x + 2^{m-1} \bmod{M}, y' = y + 2^{n-1} \bmod{N}$ such that $x' = x'_0 + x'_1 \bmod\ M$ and $y' = y'_0 + y'_1 \bmod\ N$.
Our protocol for $\fsmult{m,n}$ builds on the following equations over $\bbZ$: 
\begin{align*}
    \sigint(x) \cdot \sigint(y) =&~(x' - 2^{\sint - 1}) \cdot (y' - 2^{\lint - 1}) \text{\quad from Eq.~\ref{eq:sext}} \\
    =&~ x'\cdot y' - 2^{\sint - 1} y'- 2^{\lint - 1} x' + 2^{\sint + \lint - 2} \\
    =&~ x'\cdot y' - 2^{\sint - 1} (y'_0 + y'_1 - 2^{\lint} w_{y'}) \\
    &~ - 2^{\lint - 1} (x'_0 + x'_1 - 2^{\sint} w_{x'}) + 2^{\sint + \lint - 2},
\end{align*}
where $w_{x'}=\wrap(x'_0, x'_1, M), w_{y'} = \wrap(y'_0, y'_1, N)$.

In the protocol, parties can compute the shares of $x', y'$ locally. All terms in the final expression can be computed and added locally except  $z_1 = x' y'$ and $z_2 = 2^{\ell-1}  (w_{x'} + w_{y'})$.
Since the final expression needs to be computed  $\bmod~\intx$, we can compute shares of $z_1$ in $\bbZ_{\intx}$ using a call to $\protumult{m,n}$.
We piggyback the computation of boolean shares of $w_{x'}$ and $w_{y'}$ on $\protumult{m,n}$, which already computes them in steps~\ref{umult-wrap-one}\&\ref{umult-wrap-two}.
Note that $2^{\ell-1} w_{x'} = 2^{\ell-1} (\share{w_{x'}}{B}{0} + \share{w_{x'}}{B}{1} - 2\share{w_{x'}}{B}{0}\share{w_{x'}}{B}{1})$ and taking a $\bmod\ L$ gets rid of the last term.
Hence, $2^{\ell-1} (\share{w_{x'}}{B}{} + \share{w_{y'}}{B}{})$ are correct arithmetic shares of $z_2$ in $\bbZ_{\intx}$.
Thus, we can do signed multiplication with a single call to $\protumult{m,n}$ and no additional cost.
\par
We also consider the signed-multiplication functionality $\fsmult{m,n,\ell}$, where the output bitwidth $\ell < m + n$.
The above discussion on signed-multiplication holds in this case as well, and thus, $\protsmult{m,n,\ell}$ has the same cost as $\protumult{m,n,\ell}$. \\

\noindent{\em Matrix Multiplication and Convolutions.} 
Two commonly used operations in machine learning are 
matrix multiplications and convolutions  that build on element-wise multiplications. Consider matrix multiplication of $A \in \bbZ_{M}^{d_1 \times d_2}$ and $B \in \bbZ_{N}^{d_2 \times d_3}$, where we would like to use our protocol for $\fumult{m,n}$. 
Now, each element in the output product matrix is a result of $d_2$ multiplications and $d_2-1$ additions and even when the result of multiplication is stored in the larger ring $\bbZ_L$, $\ell = m+n$, the value can overflow due to additions. One way to avoid this overflow is to extend the result of element-wise products by $e = \lceil \log{d_2} \rceil$ bits and then do the additions. 
However, this method is quite expensive as the number of extensions needed would be $d_1d_2d_3$. 
We significantly reduce this cost as follows: Since the cost of $\fcross{}$ depends on the smaller of the two bitwidths, we extend the values in the matrix of larger bitwidth by $e$ bits. Then we perform the matrix multiplications into $\bbZ_{2^{m+n+e}}$, ensuring that there are no overflows.
Moreover, similar to the OT-based matrix multiplication from prior works \cite{secureml,cryptflow2}, we also exploit the multi-use of input matrix elements to optimize the cost of computing (matrix) cross-terms in our protocol.
Our protocol has communication complexity roughly $\secpar\left(3d_1d_2\left(m+2\right) + d_2d_3\left(n+2\right)\right) + d_1d_2d_3 \left((2m+4)(n+e) + m^2 + 5m\right)$ bits for $m \leq n$ ignoring lower order terms.
We describe our protocol formally in \appendixref{matrix-mult} along with exact communication complexity.
Above ideas easily extend to computing convolutions as well. \\

\noindent{\em Multiply and Truncate.}
In most of our protocols, we first invoke $\fsmult{m,n,\ell}$ followed by $\ftr{\ell,s}$, where $\ell \leq m+n$. 
Hence, for ease of exposition, we define the functionality  $\fsmultTR{m,n,\ell,s}$ for signed multiplication and truncate-reduce that takes $\share{x}{\sint}{}$ and $\share{y}{\lint}{}$ as input and returns $\share{z'}{\ell-s}{}$ such that $z = \sigint(x) \umult{\ell} \sigint(y)$ and $z' = \TR{z}{s}$. %

\subsection{Digit Decomposition and MSNZB}
\label{sec:digdec-short}

We consider the functionality $\fdigdec{\ell, \{d_i\}_{i \in [c]}}$ that decomposes an $\ell$-bit number into $c$ sub-strings or digits of lengths $\{d_i\}$. 
More formally, $\fdigdec{\ell,  \{d_i\}_{i \in [c]}}$ takes $\share{x}{\ell}{}$ as input and outputs 
$\share{z_{c-1}}{{d_{c-1}}}{}, \ldots, \share{z_0}{{d_0}}{}$ 
such that $x = z_{c-1} \concat \ldots \concat z_0$.

For an $\ell$-bit integer $x$, $\msnzb(x)$ refers to the index of the most significant non-zero-bit. That is, $\msnzb(x) = k \in [\ell]$, if $x_k = 1$ and $x_j = 0$ for all $j > k$.
Consider the functionality $\fmsnzb{\ell}$ that takes as input $\share{x}{\ell}{}$ and outputs $\{\share{z_i}{B}{}\}_{i \in [\ell]}$ such that $z_i = 1$ if $\msnzb(x) = i$ and $0$ otherwise.

We describe the protocols for $\fdigdec{\ell,\{d_i\}_{i \in [c]}}$ and $\fmsnzb{\ell}$ in  \appendixref{digit-decomp} and \ref{app:msnzb}, respectively.
\subsection{MSB-to-Wrap Optimization} 
\label{sec:msb-to-wrap-opt}

Our protocols above for extension, truncation and multiplication make use of the following step: Parties $P_0, P_1$ hold $\share{x}{\ell}{}$ and compute $\share{w}{B}{}$, where $w = \wrap(\share{x}{\ell}{0}, \share{x}{\ell}{1}, L)$. This is either computed through an explicit call to $\fwrap{\ell}$ (e.g., extension and multiplication) or computed via wrap of lower and upper bits (e.g., truncation). We show that shares of $w$ can be computed with much less communication and rounds if the parties either know the $m_x = \msb(x)$ in the clear or shared form. The MSB refers to the most significant bit of a number. In our math library implementations in \sectionref{mathlib-protocols}, this condition is true for almost all invocations. For instance, in exponential, when multiplying the values from multiple LUTs, we know that all operands are positive, i.e., MSB of all inputs to multiplication is $0$.
We call this optimization {\em MSB-to-Wrap} and the idea is as follows: We can write $w = ((1 \xor m_x) \wedge (m_0 \oplus m_1)) \oplus (m_0 \wedge m_1)$, where $m_b = \msb(\share{x}{\ell}{b})$ for $b \in \zo$. With this, given shares of $m_x$, boolean shares of $w$ can be computed using a single call to $\kkot{4}{1}$, i.e., $2\secpar+4$ bits of communication and $2$ rounds. 
Also, when $m_x$ is publicly known, this can be computed using $\kkot{2}{1}$, i.e., $\secpar+2$ bits.
The cost of our protocols with above optimization are provided in \tableref{comm-costs}.

\section{Math Library Functions}
\label{sec:mathlib-protocols}
In this section, we provide our cleartext implementations for math functions exponential, sigmoid, tan hyperbolic (tanh), and reciprocal square root as well as the protocols for the same. Note that these functions are impossible to implement exactly using finite-bit arithmetic, and hence, our implementations realize them approximately (\sectionref{validate}). %
Below, we use the notation from \sectionref{notation} and \sectionref{crypto-bb}.
For a mathematical function $f$, we consider the functionality $\fmath{m, s, n, s'}$ that takes as input the shares $\share{x}{m}{}$ and outputs $\share{y}{n}{}$ such that $\flt{y}{n}{s'} \approx  f(\flt{x}{m}{s})$.

Our math function implementations  rely on functions discussed in \sectionref{bb}, and we recall some of them here.
We denote signed-extension of an $m$-value to an $n$-value by  $\SX{x}{m}{n}$ with $n > m$.
Next, we denote truncate-and-reduce by $s$-bits using $\TR{x}{s}$ that takes a value $x$ of, say, $\ell$-bits, drops lower $s$ bits and returns the corresponding ($\ell-s$)-bit value.
Finally, we use a signed multiplication where the operands and the output can have unequal bitwidths.
It is denoted by $x \umult{\ell} y$, where $x$ and $y$ are, say, $m$ and $n$-bit integers, respectively, and the output of multiplication is $z = \sigint(x) \cdot \sigint(y) \bmod{L}$.

\subsection{Exponential}
\label{sec:exp}

Consider the math functionality $\fexp{m,s,n,s'}$ with $\mathsf{rExp}(z) = e^{-z}$, $z \in \mathbb{R}^+$ described in~\figureref{exp}. Intuitively, the correctness of this functionality, i.e., $\flt{y}{n}{s'} \approx \rexp(\flt{x}{m}{s})$, relies on 
$\rexp(\flt{x}{m}{s}) = \rexp(2^{d(k-1)-s}x_{k-1}) \cdot \ldots \cdot \rexp(2^{-s}x_0)$.
Each $\rexp$ call on the RHS can be computed approximately using a lookup table $L$ of size $2^d$ with $s'+2$ bit entries of scale $s'$.
Since the entries of the LUTs are between $0$ and $1$ with scale $s'$, it is sufficient to have a bitwidth of $s'+2$.
 For instance, when $m = n = 16$, $d = 8$, and $s'=14$ we use two LUTs where first maps the upper $8$ bits of $x$ and second maps the lower $8$ bits of $x$. Final output is computed by multiplying the two $16$-bit looked up values from the two LUTs into a $32$-bit number followed by an appropriate truncate and reduce operation to get $16$-bit $y$ with scale $14$.
 We formally verify that for $m,s,n,s'$ used in our evaluation, our choice of $d$ ensures precise results in \sectionref{validate}.
 
 The protocol for this functionality can be built easily relying on the protocols described in \sectionref{bb}. \stepref{exp-one} can be implemented by a call to the digit decomposition functionality, $\fdigdec{}$.
 The LUTs in \stepref{exp-two} can be looked up using $\fLUT{}$ (\sectionref{two-pc-prelims}). These $s'+2$-bit values are multiplied using a tree-based multiplication using $\fsmultTR{s'+2, s'+2,2s'+2, s'}$ to get an $s'+2$-bit number with scale $s'$ in \stepref{exp-three}. 
 Finally, \stepref{exp-four} extends $g$ to an $n$-bit value using $\fsxt{s'+2, n}$.
\tableref{recsqrt-exp-micro} gives our concrete numbers and compares with prior work.

\begin{tffbox}
\begin{mdframed}
\begin{center}
{\bf Functionality} $\fexp{m,s,n,s'}(\share{x}{m}{})$
\end{center}
{\small

\begin{enumerate}
       \item Let $x = x_{k-1} \concat \ldots \concat x_0$, $x_i \in \zo^d$, $i \in [k]$, $dk = m$. \label{step:exp-one}
       \item For $i \in [k]$, let $L_i : \zo^d \rightarrow \bbZ_{2^{s'+2}}$ s.t. $L_i(j)= \fixed{\rexp(2^{di-s}j)}{s'+2}{s'}$.  \label{step:exp-two}
        \item Compute $g = L_{k-1}[x_{k-1}] \umult{} \ldots \umult{} L_0[x_0]$, $g$ has bitwidth $s'+2$ and scale $s'$. \label{step:exp-three}
       \item %
       Return $\share{y}{n}{}$ for $y = \SX{g}{s'+2}{n}$.  \label{step:exp-four}
       
\end{enumerate}
	
} 
\end{mdframed}
    \caption{\sl The functionality $\fexp{m,s,n,s'}$ for a parameter $d$.}
\label{fig:exp}
\end{tffbox}

\newcommand{\mys}{ s}

\subsection{Sigmoid and Tanh}
\label{sec:sigmoid-tanh}

Consider the math functionality $\fsigmoid{m,s,n,s'}$ where $\sigmoid(z)=\frac{1}{1+e^{-z}}$ can be written as 
\[
\sigmoid(z) = \begin{cases}
      0.5, & \text{if}\ z=0 \\
      \frac{1}{1+\rexp(z)}, & \text{if}\ z> 0\\
      \rexp(-z)\frac{1}{1+\rexp(-z)} , & \text{if}\ z< 0\\
    \end{cases}
\]
Hence, sigmoid can be built by extending the math functionality $\fh{m,s,n,s'}$ such that $h(z) = \frac{1}{1+\rexp(z)}$, $z \in \bbR^+$ described in \figureref{sigmoid-h}. This functionality calls $\fexp{}$ that we described above, followed by a call to a functionality to approximate the reciprocal that we describe next.
 
\begin{tffbox}
\begin{mdframed}
\begin{center}
{\bf Functionality} $\fh{m,s,n,s'}(\share{x}{m}{})$
\end{center}
{\small

\begin{enumerate}
\item $\share{u}{s'+2}{} \leftarrow \fexp{m,s, s'+2,s'}(\share{x}{m}{})$.
\item $\share{w}{s'+2}{} \leftarrow \frecip{s'+2, s'}(\share{2^{s'}+u}{s'+2}{})$.
\item Return $\SX{w}{s'+2}{n}$. 
  
\end{enumerate}
	
} 
\end{mdframed}
\caption{\sl The functionality $\fh{m,s,n,s'}$.}
\label{fig:sigmoid-h}
\end{tffbox}

For computing the reciprocal, we rely on the Goldschmidt's algorithm~\cite{goldschmidt} that iterates on an initial approximation~\cite{ito-lookup}. This initial approximation requires that we only compute reciprocal of values $v$ such that $1 \leq \flt{v}{\ell}{\mys} < 2$  which is true for the case of $h$ and $\sigmoid$. 

We describe the math functionality $\frecip{\ell,\mys}$ in \figureref{recip} that maps inputs $v$ with bitwidth $\ell$ and scale $\mys$ to outputs of same bitwidth and scale. 
Since $1 \leq \flt{v}{\ell}{\mys} < 2$, in \stepref{rec-step-one}, $d = 1$. 
We use the $g$ most significant bits of the fractional part to index into the LUT $L_{\mathsf{rec}}$ in \stepref{rec-lookup} whose entries are described in \cite{ito-lookup}.
The initial approximation $w$ has bitwidth $\mys+1$ and scale $\mys$. 
If the number of Goldschmidt iterations $t$ is set to $0$, then $\frecip{}$ outputs initial approximation sign extended to output bitwidth, i.e., $\SX{w}{\mys+1}{\ell}$. 
We formally verify that for $m,s,n,s'$ used in our evaluation, our choice of parameters for $\fexp{}$ and $\frecip{}$ ensures precise results for $\fsigmoid{}$ in \sectionref{validate}.

Note that this functionality crucially utilizes arithmetic over variable bitwidth and extension/truncation operations and these steps require our efficient protocols from \sectionref{bb}.
\tableref{sigmoid-micro} gives our concrete numbers and compares with prior work.

\begin{tffbox}
\begin{mdframed}
\begin{center}
{\bf Functionality} $\frecip{\ell,\mys}(\share{v}{\ell}{})$
\end{center}
{\small
Computes the initial approximation $w$ as follows~\cite{ito-lookup}:
\begin{enumerate}
\item $v = d \concat e \concat f$, $d \in \zo^{\ell-\mys}$, $e \in \zo^{g}$, $f \in \zo^{\mys-g}$. \label{step:rec-step-one}
\item $c_0 || c_1 = L_{\mathsf{rec}}(e)$, $c_0 \in \zo^{g+4}$ and $c_1 \in \zo^{2g+3}$. \label{step:rec-lookup}
\item $c_2 = \SX{(c_0 \umult{\mys+4} f)}{\mys+4}{\mys+g+4}$.
\item $w' = 2^{\mys - g +1} \umult{\mys+g+4} c_1 - c_2$, $w = \TR{w'}{g+3}$.
\end{enumerate}

Goldschmidt's method for $t$ iterations.
\begin{enumerate}
    \item $p_1 = 2^{\mys} - \TR{v \umult{2\mys+2} w}{\mys}$.
    \item $q_1 = 2^{\mys} + p_1, a_1 = q_1$.
    \item For $i \in \{ 2, \ldots, t \}$ do
    \begin{enumerate}
        \item $a_i = \TR{a_{i-1} \umult{2\mys+2} q_{i-1}}{\mys}$.
        \item $p_i = \TR{p_{i-1} \umult{2\mys+2} p_{i-1}}{\mys}$.
        \item $q_i = 2^{\mys} + p_i$.
    \end{enumerate}
\item Return $\SX{a_{t}}{\mys+2}{\ell}$.
\end{enumerate}
	
} 
\end{mdframed}
\caption{\sl The functionality $\frecip{\ell,\mys}$ for a parameters $g,t$.}
\label{fig:recip}
\end{tffbox}

\noindent{\em Tanh.} The math functionality $\ftanh{m,s,n,s'}$ where $\mytanh(z)=\frac{e^z - e^{-z}}{e^z+e^{-z}} = 2 \cdot \sigmoid(2z) -1$ can be realized using $\fsigmoid{}$.

\subsection{Reciprocal of Square Root} \label{sec:inverse-sqrt}
In ML, reciprocal square root is typically used to scale down vectors $\vec{u}$ of large magnitude to unit vectors by dividing each entry of the vector with $\frac{1}{\sqrt{u^Tu}}$.
The reciprocal square root function maps $x$ to $\frac{1}{\sqrt{x}}$, for $x > 0$.
If $x$ is small then to avoid divide-by-zero errors
a small public constant $\epsilon$ is added to $x$ and $\frac{1}{\sqrt{x+\epsilon}}$ is computed instead.
Hence, we present our mathematical functionality $\frsqrt{\ell,s,\ell, s'}$  in \figureref{rsqrt} for the math function $\rsqrt(z)=\frac{1}{\sqrt{z}}$ where $z\geq \epsilon$. 

This functionality follows a similar template of first computing an initial approximation for reciprocal square root followed by Goldschmidt's iterations. The initial approximation~\footnote{
Although we would have liked to use the initial approximation provided by \cite{ito-lookup}, there seems to be some typographical errors in the published equations and we are unable to correct them.
}
requires $1 \leq x < 2$, and hence, first we perform a range reduction to map arbitrary $x$ of the form $y.z$ to $x'$ of the form $1.z'$ that satisfies this constraint.  This requires computing the most significant non-zero bit (MSNZB) of $x$ (\stepref{rsqrt-msnzb}). Note that $\msnzb(x) = k \in [\ell]$ if $x_k = 1$ and all $x_i = 0$ for all $i > k$.
The normalized value $x'$ has bitwidth $\ell$ and scale $\ell-2$.
Next, we use $g$ most significant bits of $z'$, i.e., $e$ and the parity of $k-s$, i.e., $B$,  to compute the initial approximation via a lookup table
 $L_{\rsqrt}$ whose entries are as follows: 
$$L_{\rsqrt}(e\concat B) = \fixed{\frac{1}{\sqrt{(B+1)(1+\uflt{e}{g}{g})}}}{g+4}{g+2} $$

\begin{tffbox}
\begin{mdframed}
\begin{center}
{\bf Functionality} $\frsqrt{\ell,s,\ell, s'}(\share{x}{\ell}{})$
\end{center}
{\small
Normalizes $x$ to   $x'$ as follows:
\begin{enumerate}
	\item $k = \msnzb(x) \in [\ell]$. \label{step:rsqrt-msnzb}
	\item $A = 2^{\ell-2-k}$, $B = (s-k) \bmod\ 2$.
    \item $C = 2^{\lceil\frac{s-k}{2}\rceil+\lfloor\frac{\ell-s-1}{2}\rfloor}$.
    \item $x' = x \umult{\ell} A$.
\end{enumerate}
Computes the initial approximation $w$ as follows:
\begin{enumerate}
    \item ${x' = d \concat e \concat f, d \in \zo^{2}, e \in \zo^{g}, f \in \zo^{\ell-2-g}}$.
    \item $w = L_{\rsqrt}(e \concat B)$, $w \in \zo^{g+4}$.
\end{enumerate}

Goldschmidt's method for $t$ iterations:
\begin{enumerate}
    \item $x'' = \TR{x'}{\ell-3-s'}, ~q_0 = B ~ ? ~ x'' : \TR{x'}{1}$.
    \item $a_0 = 2^{s' - g - 2} \umult{s' + 2} w,~ p_0 = a_0$.
    \item For $i \in \{ 1, \ldots, t \}$ do
    \begin{enumerate}
        \item $Y_{i} = \TR{p_{i-1} \umult{2s'+2} p_{i-1}}{s'}$.
        \item $q_{i} = \TR{q_{i-1} \umult{2s'+2} Y_i}{s'} $.
        \item $p_{i} = 3 \cdot 2^{s'-1} - ({q_{i}}\ggA{1})$.
        \item $a_{i} = \TR{a_{i-1} \umult{2s' + 2} p_i}{s'}$.
    \end{enumerate}
\end{enumerate}
Uses reciprocal square root of $x'$ to compute the same for $x$:
\begin{enumerate}
    \item Return $\TR{a_{t} \umult{\ell/2 + s' + 3} C}{\big\lfloor\frac{\ell-s-1}{2}\big\rfloor} \bmod{\intx}$.
\end{enumerate}
} 
\end{mdframed}
\caption{\sl The functionality $\frsqrt{\ell,s,s'}$ for parameters $g,t$.}
\label{fig:rsqrt}
\end{tffbox}

We formally verify that for $\ell,s,s'$ in our evaluation, our choice of $g,t$  ensures precise results for $\frsqrt{}$ (\sectionref{validate}). 

We build a protocol for $\frsqrt{}$ as follows: We consider the functionality $\fmsnzb{}$ that outputs the shares of one-hot encoding of $\msnzb(x)$ and give a protocol for the same in \appendixref{msnzb}. It is easy to compute the terms $A, B, C$ using dot-products of this one-hot vector with publicly known vectors. For our initial approximation, we rely on protocols for $\fdigdec{}$ and $\fLUT{}$. The Goldschmidt's iterations crucially utilize arithmetic over variable bitwidth and truncation operations and each of these steps require our efficient protocols from \sectionref{bb}.
\tableref{recsqrt-exp-micro} gives our concrete numbers and compares with prior work.

\subsection{Formal verification of our Math functionalities} \label{sec:validate}
It is desirable for math libraries to have a formal proof of
correctness about their purported numerical precision.
Such a proof establishes that for all possible inputs, the ULP error
(\sectionref{prelims}) between the math implementation and the exact
real result is small.
For small bitwidths (e.g. $\leq 32$) that are used in ML
(\sectionref{eval-cases}), it is tractable to prove these bounds
on ULP error using {\em exhaustive testing}, whereas
for 64-bit floating-point or 64-bit fixed-point math libraries, these
proofs can either be interactive~\cite{harrison,inrialog} or fully
automatic~\cite{rosa,fptaylor,popl18}.
Since our focus is on math libraries for ML, we choose the exhaustive
testing approach for our math library, specifically, we
1) run our implementations on all possible inputs,
2) compare the ULP error between each output and the infinite
precision real result, and
3) report the maximum observed ULP error as the bound.
For step 2, we need the ability to compute math functions to
arbitrary degrees of precision -- this is offered by the
GNU MPFR library~\cite{mpfr}.

We prove ULP error bounds for bitwidth $16$ (\sectionref{eval-cases}) and appropriate input/output
scales, $s_x$ and $s_y$, and choose parameters $d$, $g$, and $t$
accordingly to ensure high precision.
Note that given a bitwidth $\ell$, a proof via exhaustive testing
requires $2^{\ell}$ tests.
For exponential, we set $d = 8$ and prove that $\forall s_x,s_y\in [8,14]$,
the maximum ULP error is 3.
For sigmoid and tanh, we set $d = 8$, $g = \lceil \frac{s_y - 2}{2}
\rceil$ and $t = 0$, and prove
that $\forall s_x,s_y\in[8,14]$ the maximum ULP error is $3$ for sigmoid
and $4$ for tanh.
For reciprocal square root, we choose inputs $x\geq \epsilon$ where
$\epsilon=0.1$, and set $g = \lceil \frac{s_y}{2} \rceil$ and $t =
1$. We prove that $\forall s_x,s_y\in [4,13]$, the maximum ULP error
is $4$.

Thus, using exhaustive testing, we prove that our math
implementations are precise for chosen parameters and
provide standard precision guarantees that are expected from math
libraries viz. ULP error $<5$; Intel's SVML~\cite{svml} also provides math implementations with 4 ULP error.
We use the same parameter setting described above for the empirical evaluation.

\section{Evaluation}
\label{sec:eval}
In this section, we empirically compare our protocols for math
functions with prior works  and describe the results of our ML case
studies.
The closest work to ours is MiniONN~\cite{minionn}, the only prior
work on secure inference that has been evaluated on an RNN.
MiniONN proposes a recipe to obtain piecewise linear approximations to
sigmoid/tanh that are then evaluated using its protocols.
Our secure
implementations of sigmoid  are an order of magnitude better in
communication (\tableref{sigmoid-micro}). %
Note that no prior work on 2-party secure inference (including MiniONN) provides secure implementations of exponentiation and reciprocal square root; we evaluate them in \tableref{recsqrt-exp-micro}.
General-purpose MPC frameworks like MP-SPDZ~\cite{mpspdz} also provide semi-honest 2PC implementations of math functions~\cite{mpspdz-code} that are compatible with the standard (power-of-2 ring-based) fixed-point representation.
However, the communication of our protocols is up to two orders of magnitude lower. 
Alternatives that use  representations 
such as field-based representations or floating-point also suffer from high communication overheads. 

Next, we evaluate our library \tool for DNN inference on end-to-end ML models. 
First, we evaluate \tool on models with math functions considered by priors works~\cite{minionn,deepsecure}.
Since they evaluate sigmoid and tanh using generic 2PC protocols, \tool has an order of magnitude less communication (\tableref{prior-dnn}). 
Next, we evaluate \tool on RNNs for sports training and audio keyword
spotting that use GRU cells, which are composed of sigmoid and tanh
operators. There are two ways to securely evaluate our
math functionalities, with our 2PC protocols and with generic 2PC
protocols for mixed arithmetic and boolean
compute~\cite{aby,aby2,ezpc,hycc}.
We evaluate both and observe that \tool communicates over $500\times$ less data for both the RNNs (\tableref{rnns}). 
Finally, we evaluate \tool on a recent model architecture that combines CNN operators and RNN operators to find the human heads in images with state-of-the-art accuracy~\cite{rnnpool}.  
We provide the first secure implementation for this complex model; its
secure implementation requires all the protocols described in this
paper including reciprocal square root and takes less than 7 minutes
on our evaluation set up:

\noindent {\em System Details.} We use a set up 
where the 2 machines are connected via a 377 MBps LAN network with 0.8
ms RTT. Both the machines have commodity hardware with a 4-core
3.7 GHz Xeon processor and 16 GBs of RAM.

\noindent {\em Implementation Details.} The users of \tool express
their DNNs as a combination of calls to \tool's C++ library
functions. These functions include matrix multiplication,
convolutions, MBConv blocks, L2 Normalization, batch normalization,
broadcasting; pointwise operators
like sigmoid, tanh, exponential, reciprocal square root, 
matrix addition, Hadamard product; comparison-based operators like  argmax, maxpool, ReLU, and ReLU6. 
The last four functions use protocols from~\cite{cryptflow2} and the rest use
our building blocks.
The  library functions take
scales as arguments and are templated on the bitwidths. The \tool
library is implemented using 28K lines of C++. We statically generate
36 LUTs that consume additional 35K LOC.

\subsection{Microbenchmarks}
\label{sec:micro}

\paragraph{Sigmoid} In \tableref{sigmoid-micro}, we compare our
protocol 
with prior work for generating sigmoid output with 12-bits of precision (i.e., scale 12).
We report absolute numbers for time taken and communication for both
our protocols and prior work, as
well as improvement factor of our protocols in parentheses. We follow
this pattern for all the tables in this section.
We focus on sigmoid as the numbers for tanh are similar.
One sigmoid evaluation with our protocols incurs less than 5KB of
communication and produces precise results with at most 3 ULPs 
error.
In ML, sigmoid is usually computed pointwise over all the entries in a
tensor.  Hence, one needs to compute sigmoid of a large number of
instances when dealing with realistic ML benchmarks.
Although the communication to compute $n$ sigmoid instances grows
linearly with $n$, empirically we have observed that the time taken or
the latency  grows sub-linearly with $n$ (columns 2 to 5 of
\tableref{sigmoid-micro}), which helps our implementations to
scale well to large tensors (\sectionref{eval-cases}).
The cost of rounds amortizes better for large tensors resulting in the
sub-linear growth in latency.

As a baseline, we consider the recipe of MiniONN that approximates
math functions with piecewise linear approximations and provides
protocols to evaluate these splines. More precise approximations
require more number of pieces. To get an ULP error below 5, MiniONN
needs a 48-way spline which provides poor performance when evaluated
securely because of a $70\times$ communication overhead.

For the RNN benchmark that MiniONN considers
(\sectionref{priordnns}), the precision offered by the 48-piece
spline is an overkill and a 12-piece spline  suffices to maintain the
cross entropy loss. 
Although this 12-piece spline is more efficient than 48-piece spline, its performance is still much worse than our protocols and incurs a $19\times$ communication overhead.
Furthermore, this 12-piece spline incurs an error of 104 ULPs. Hence,
our implementations are superior in both precision and performance.
While a 12-piece spline suffices for this benchmark, MiniONN remarks that other benchmarks need splines with more number of pieces that are even more expensive to compute.
Because our implementations are guaranteed to be numerically precise,
they can be used as-is with no loss in model accuracy
(\sectionref{eval-cases}).

DeepSecure~\cite{deepsecure} uses garbled circuits (GC) to evaluate DNNs that use sigmoid and tanh activations. We checked with the authors of DeepSecure and the circuits for math functions are not available. Hence, we cannot compute the ULP errors of their implementations.
However, DeepSecure reports the number of non-XOR gates that can be used for performance estimates. 
We used state-of-the-art for GC implementation, i.e., EMP-Toolkit~\cite{emp-toolkit,GuoKW020,GuoKWW020}, to obtain these performance estimates that are better than the performance reported by DeepSecure.
The communication of our protocols is $25\times$ lower ($4^{\mathrm{th}}$ row of \tableref{sigmoid-micro}).

MP-SPDZ~\cite{mpspdz}, a general-purpose MPC framework, provides 2 baseline sigmoid implementations for 2PC~\cite{mpspdz-code}: Poly-based, which uses a range reduction and Taylor series polynomials to compute exponential followed by division, and PL-based, which is a built-in piecewise linear spline. The former implementation incurs error comparable to us but communicates $201\times$ more, while the latter is more than an order of magnitude inferior in precision and communication ($5^{\mathrm{th}}$ and $6^{\mathrm{th}}$ row of \tableref{sigmoid-micro}).
\begin{table}[]
\centering
\renewcommand*{\arraystretch}{1.1}
\begin{tabular}{|c|c|c|c|c|c|c|}
\hline
\multirow{3}{*}{Technique}                                            & \multicolumn{4}{c|}{Total Time for \#Instances (in sec)}                                                                                                                                                                                 & \multirow{3}{*}{\begin{tabular}[c]{@{}c@{}}Comm./\\ Instance\\ (in KB)\end{tabular}} & \multirow{3}{*}{\begin{tabular}[c]{@{}c@{}}Max \\ ULP\\ Err.\end{tabular}} \\ \cline{2-5}
                                                                      & \multirow{2}{*}{$10^2$}                                    & \multirow{2}{*}{$10^3$}                                   & \multirow{2}{*}{$10^4$}                                   & \multirow{2}{*}{$10^5$}                                   &                                                                                         &                                                                             \\
                                                                      &                                                         &                                                         &                                                          &                                                           &                                                                                         &                                                                             \\ \hline
\hline
Our Work                                                             & $0.08$                                                    & $0.10$                                                    & $0.25$                                                     & $1.58$                                                      & $4.88$                                                                                    & $3$                                                                           \\ \hline
\begin{tabular}[c]{@{}c@{}}MiniONN\\$48$-piece \end{tabular}    & \begin{tabular}[c]{@{}c@{}}$0.20$\\ ($2.5$x)\end{tabular}  & \begin{tabular}[c]{@{}c@{}}$1.94$\\ ($19.4$x)\end{tabular}  & \begin{tabular}[c]{@{}c@{}}$18.85$\\ ($75$x)\end{tabular}  & \begin{tabular}[c]{@{}c@{}}$182.2$\\ ($115$x)\end{tabular}  & \begin{tabular}[c]{@{}c@{}}$341.03$\\ ($70$x)\end{tabular}                                   & $4$                                                                         \\ \hline
\hline
\begin{tabular}[c]{@{}c@{}}MiniONN\\$12$-piece \end{tabular}                                                           & \begin{tabular}[c]{@{}c@{}}$0.06$\\ ($0.8$x)\end{tabular}  & \begin{tabular}[c]{@{}c@{}}$0.54$\\ ($5.4$x)\end{tabular}  & \begin{tabular}[c]{@{}c@{}}$5.24$\\ ($21$x)\end{tabular}  & \begin{tabular}[c]{@{}c@{}}$53.84$\\ ($34$x)\end{tabular}  & \begin{tabular}[c]{@{}c@{}}$93.36$\\ ($19.1$x)\end{tabular}                                   & $104$                                                                         \\ \hline
\begin{tabular}[c]{@{}c@{}}Deep-\\Secure \end{tabular}                                                            & \begin{tabular}[c]{@{}c@{}}$0.16$\\ ($2$x)\end{tabular}  & \begin{tabular}[c]{@{}c@{}}$0.84$\\ ($8.4$x)\end{tabular}    & \begin{tabular}[c]{@{}c@{}}$8.1$\\ ($32$x)\end{tabular}  & \begin{tabular}[c]{@{}c@{}}$141.3$\\ ($89$x)\end{tabular} & \begin{tabular}[c]{@{}c@{}}$124.65$\\ ($25$x)\end{tabular}                                & NA                                                                           \\ \hline
\hline
\begin{tabular}[c]{@{}c@{}}MP-SPDZ\\Ring Poly \end{tabular}    & \begin{tabular}[c]{@{}c@{}}$0.75$\\ ($9.4$x)\end{tabular}  & \begin{tabular}[c]{@{}c@{}}$1.72$\\ ($17.2$x)\end{tabular}  & \begin{tabular}[c]{@{}c@{}}$14.88$\\ ($59.5$x)\end{tabular} & \begin{tabular}[c]{@{}c@{}}$140.6$\\ ($89$x)\end{tabular}  & \begin{tabular}[c]{@{}c@{}}$981.11$\\ ($201$x)\end{tabular}                               & $2$                                                                          \\ \hline
\begin{tabular}[c]{@{}c@{}}MP-SPDZ\\Ring PL \end{tabular}  & \begin{tabular}[c]{@{}c@{}}$0.27$\\ ($3.4$x)\end{tabular}  & \begin{tabular}[c]{@{}c@{}}$0.28$\\ ($2.8$x)\end{tabular}     & \begin{tabular}[c]{@{}c@{}}$1.32$\\ ($5.3$x)\end{tabular}   & \begin{tabular}[c]{@{}c@{}}$12.34$\\ ($7.8$x)\end{tabular}   & \begin{tabular}[c]{@{}c@{}}$76.42$\\ ($15.7$x)\end{tabular}                                & $266$                                                                         \\ \hline
\hline
\begin{tabular}[c]{@{}c@{}}MP-SPDZ\\ Field Poly \end{tabular}   & \begin{tabular}[c]{@{}c@{}}$0.91$\\ ($11.4$x)\end{tabular}  & \begin{tabular}[c]{@{}c@{}}$1.91$\\ ($19.1$x)\end{tabular} & \begin{tabular}[c]{@{}c@{}}$16.51$\\ ($66$x)\end{tabular} & \begin{tabular}[c]{@{}c@{}}$127$\\ ($80$x)\end{tabular} & \begin{tabular}[c]{@{}c@{}}$228.63$\\ ($46.9$x)\end{tabular}                               & $2$                                                                          \\ \hline
\begin{tabular}[c]{@{}c@{}}MP-SPDZ\\ Field PL \end{tabular} & \begin{tabular}[c]{@{}c@{}}$0.52$\\ ($6.5$x)\end{tabular}  & \begin{tabular}[c]{@{}c@{}}$0.47$\\ ($4.7$x)\end{tabular}  & \begin{tabular}[c]{@{}c@{}}$1.79$\\ ($7.2$x)\end{tabular}   & \begin{tabular}[c]{@{}c@{}}$14.23$\\ ($9$x)\end{tabular}   & \begin{tabular}[c]{@{}c@{}}$27.52$\\ ($5.6$x)\end{tabular}                                 & $266$                                                                         \\ \hline
\end{tabular}
    \caption{Comparison with prior works on  sigmoid with varying number of instances.}
\label{tab:sigmoid-micro}
\end{table}

\begin{table}[]
\centering
\renewcommand*{\arraystretch}{1.1}
\begin{tabular}{|c|c|c|c|c|c|c|}
\hline
\multirow{3}{*}{Technique} &
  \multicolumn{4}{c|}{Total Time for \#Instances (in sec)} &
  \multirow{3}{*}{\begin{tabular}[c]{@{}c@{}}Comm./\\ Instance\\ (in KB)\end{tabular}} &
  \multirow{3}{*}{\begin{tabular}[c]{@{}c@{}}Max \\ ULP\\ Error\end{tabular}} \\ \cline{2-5}
          & \multirow{2}{*}{$10^2$} & \multirow{2}{*}{$10^3$} & \multirow{2}{*}{$10^4$} & \multirow{2}{*}{$10^5$} &      &   \\
          &                      &                       &                        &                         &      &   \\ \hline
\hline
\multicolumn{7}{|c|}{Exponentiation} \\ \hline
Our Work & 0.03 & 0.04 & 0.15 & 1.00 & 2.12 & $3$ \\ \hline
MP-SPDZ &
  \begin{tabular}[c]{@{}c@{}}$0.34$\\ ($11.3$x)\end{tabular} &
  \begin{tabular}[c]{@{}c@{}}$0.56$\\ ($14$x)\end{tabular} &
  \begin{tabular}[c]{@{}c@{}}$3.90$\\ ($26$x)\end{tabular} &
  \begin{tabular}[c]{@{}c@{}}$35.95$\\ ($35.9$x)\end{tabular} &
  \begin{tabular}[c]{@{}c@{}}$254.95$\\ ($120$x)\end{tabular} &
  $2$\\ \hline \hline
\multicolumn{7}{|c|}{Reciprocal Square Root} \\ \hline
Our Work & $0.13$                 & $0.13$                  & $0.30$                   & $1.84$                    & $6$ & $4$ \\ \hline
MP-SPDZ &
  \begin{tabular}[c]{@{}c@{}}$0.94$\\ ($7.2$x)\end{tabular} &
  \begin{tabular}[c]{@{}c@{}}$3.90$\\ ($30$x)\end{tabular} &
  \begin{tabular}[c]{@{}c@{}}$35.87$\\ ($120$x)\end{tabular} &
  \begin{tabular}[c]{@{}c@{}}$338.9$\\ ($184$x)\end{tabular} &
  \begin{tabular}[c]{@{}c@{}}$2535$\\ ($423$x)\end{tabular} &
  $8$ \\ \hline
\end{tabular}
    \caption{Comparison with (power-of-2) ring-based MP-SPDZ protocols with varying number of instances.}
\label{tab:recsqrt-exp-micro}
\end{table}

While we focus on power-of-2 rings, there are other works on secure implementations of sigmoid that use field-based or floating point representations. 
Field-based protocols perform poorly for non-linear computations like truncation and comparisons, which are abundant in fixed-point representations of DNNs~\cite{gazelle,delphi,cryptflow2}. Similarly, it is well-known that the protocols over floating-point are much slower than fixed-point~\cite{Catrina19,cryptflow}. 
Nonetheless, for completeness, we compare against the 
 state-of-the-art field-based implementations in MP-SPDZ~\cite{mpspdz,mpspdz-code} and they perform worse ($7^{\mathrm{th}}$ and $8^{\mathrm{th}}$ rows of \tableref{sigmoid-micro}). 
 We also compare with floating-point implementations of math functions provided by ABY~\cite{abyfloat} and EMP-Toolkit~\cite{emp-toolkit}; our protocols are at least $90\times$ better in communication per instance and $97\times$ better in runtime (for $10^5$ instances).

Finally, SecureML~\cite{secureml} and ABY2.0~\cite{aby2} use a 3-piece linear spline to approximate sigmoid. This simple implementation has a whopping error of 1547 ULPs and tanks the accuracy of our RNN benchmarks.
For instance, it leads to a tremendous drop in accuracy of the Google-30 network from 84.4\% (with our sigmoid implementation) to 60.95\%.
The insufficiency of this approximation has also been noted by~\cite{minionn} where it caused the cross-entropy loss to diverge to infinity. Hence, this crude approximation is usable only in restricted contexts and is unsuitable for generic math libraries, which is our aim~here.

\paragraph{Exponential and reciprocal square-root}
\tableref{recsqrt-exp-micro} shows the comparison of our
exponentiation and reciprocal
square-root protocols with power-of-2 ring based protocols in MP-SPDZ framework (for scale $12$). It has native support for exponentiation.
We implement reciprocal square root in MP-SPDZ by calling its built-in functions for square root and reciprocal.
As the table shows, our protocols are orders of
magnitude better, both in terms of time-taken and communication, and
provide better or comparable ULP errors.

\subsection{Prior DNNs}
\label{sec:priordnns}
In \tableref{prior-dnn}, we evaluate our protocols on
benchmarks with math functions from MiniONN~\cite{minionn} and DeepSecure~\cite{deepsecure}.
MiniONN evaluated an LSTM 
 for text data which has 2 LSTM layers each with 800 instances of sigmoid and 200 instances of tanh.
Our protocols incur an order of magnitude less communication for these instances.
We consider the largest benchmark of DeepSecure, B4, with 2 tanh layers of 2000 and 500 instances, which classifies sensor data into 19 different physical activities. To estimate the time taken by DeepSecure on our setup, we ran a circuit with the same non-XOR complexity as B4 using EMP-Toolkit~\cite{emp-toolkit} (similar to our microbenchmarks) that provides better performance than the communication and latency in~\cite{deepsecure}.
Our protocols have $87\times$ lower latency and $43\times$ lower communication.

\subsection{Case studies}
\label{sec:eval-cases}
We demonstrate the applicability of secure inference to three new
domains that no prior work has considered before: RNNs applied to time series sensor data, RNNs applied to
speech data, and combining CNNs and RNNs to identify human heads in
images.
The feasibility of our case studies crucially relies on our efficient
protocols for math functions.
Our first case study is an industrial model
(Industrial~\cite{shiftry}) which uses an RNN with GRU cells to
provide feedback on the quality of shots in a bat-and-ball game from
the data obtained from sensors deployed on the bat. Second, we
evaluate an RNN (Google-30~\cite{fastgrnn}) for keyword spotting in
the standard Google-30~\cite{Google-30} dataset that identifies simple
commands, digits, and directions from speech data obtained from
thousands of people. Third, the head detection model
(Heads~\cite{rnnpool})
combines CNNs and RNNs for the best accuracy on the SCUT Head
dataset~\cite{scut}. It uses
inverted residual blocks, or MBConv blocks~\cite{mobilenetv2}, for efficient convolutions. 
Instead of simple pooling operators like maxpool or average pool, it uses RNN-based pooling that provides  high accuracy.
We summarize the input fixed-point code of these benchmarks
below. These fixed-point C++
programs were automatically generated from high-level ML
models by~\cite{shiftry} (a compiler for embedded devices) and linked
with \tool.
All of the benchmarks use a mixture of
variables with bitwidth 8, 16, and 32 with 16 being the bitwidth used
for input and output of the math functions.
\begin{table}[]
\centering
\renewcommand*{\arraystretch}{1.1}
\begin{tabular}{|c|c|c|c|c|}
\hline
\multirow{2}{*}{\begin{tabular}[c]{@{}c@{}}Inference\\ Benchmark\end{tabular}} & \multicolumn{2}{c|}{Runtime (in sec)}                                                    & \multicolumn{2}{c|}{Comm.}                                                        \\ \cline{2-5} 
                                                                               & \multicolumn{1}{c|}{Prior}                              & \multicolumn{1}{c|}{Our Work} & \multicolumn{1}{c|}{Prior}                               & \multicolumn{1}{c|}{Our Work} \\ \hline
\hline
\begin{tabular}[c]{@{}c@{}}MiniONN LSTM\end{tabular}       & \begin{tabular}[c]{@{}c@{}}$1.1$\\ ($2.2$x)\end{tabular} & $0.48$                        & \begin{tabular}[c]{@{}c@{}}$182$ MB\\ ($19.5$x)\end{tabular} & $9.32$ MB                         \\ \hline
DeepSecure B4                                                                  & \begin{tabular}[c]{@{}c@{}}$465$\\ ($87$x)\end{tabular}  & $5.3$                            & \begin{tabular}[c]{@{}c@{}}$83.7$ GB\\ ($43$x)\end{tabular}  & $1.94$ GB                          \\ \hline
\end{tabular}
\caption{Comparison with benchmarks from MiniONN~\cite{minionn} and DeepSecure~\cite{deepsecure}. }
\label{tab:prior-dnn}
\end{table}

\begin{table}[]
\centering
\renewcommand*{\arraystretch}{1.1}
\begin{tabular}{|c|c|c|c|c|c|}
\hline
\multirow{2}{*}{\begin{tabular}[c]{@{}c@{}} Benchmark\end{tabular}} & \multirow{2}{*}{\begin{tabular}[c]{@{}c@{}} Batch\end{tabular}} & \multicolumn{2}{c|}{Runtime (sec)}                                                    & \multicolumn{2}{c|}{Comm.}                                                        \\ \cline{3-6} 
                                                                      &                                                                       & \multicolumn{1}{c|}{\cite{aby}}                              & \multicolumn{1}{c|}{\tool} & \multicolumn{1}{c|}{\cite{aby}}                               & \multicolumn{1}{c|}{\tool} \\ \hline
\hline
\multirow{3}{*}{\begin{tabular}[c]{@{}c@{}} Industrial-72\end{tabular}}   & $1$   & \begin{tabular}[c]{@{}c@{}}$68.33$\\ ($18$x) \end{tabular}         & $3.7$       & \begin{tabular}[c]{@{}c@{}}$11.84$ GB \\ ($510$x)\end{tabular}   &  $23.8$ MB             \\ \cline{2-6}
                                                                          & $128$ & \begin{tabular}[c]{@{}c@{}}$8746$$^*$\\ ($661$x) \end{tabular}       & $13.2$      & \begin{tabular}[c]{@{}c@{}}$1.47$ TB$^*$ \\ ($1451$x)\end{tabular} &  $1.04$ GB           \\ \hline        
\multirow{3}{*}{\begin{tabular}[c]{@{}c@{}} Google-30\end{tabular}}       & $1$   & \begin{tabular}[c]{@{}c@{}}$3337$ \\ ($67$x)\end{tabular}          & $49.6$      & \begin{tabular}[c]{@{}c@{}}$259$ GB \\ ($574$x)\end{tabular}     &  $0.45$ GB           \\ \cline{2-6}
                                                                          & $128$ & \begin{tabular}[c]{@{}c@{}}$4.3$x$10^5$$^*$ \\ ($3050$x)\end{tabular}    & $140$     & \begin{tabular}[c]{@{}c@{}}$32.38$ TB$^*$ \\ ($1316$x)\end{tabular}     &  $25.2$ GB           \\ \hline
Heads                                                                     & $1$   & NA                                                                      & $409.7$     & NA                                                                  & $85.5$ GB               \\ \hline
\multicolumn{6}{r}{\footnotesize{*extrapolated, the run could not be completed due to TB comm.}}\\
\end{tabular}
\caption{Secure inference on DNNs using \tool and \cite{aby}.}
\label{tab:rnns}
\end{table}

\begin{itemize}
\item {\em Industrial-72}: It contains 7 sigmoid and 7 tanh layers, with  64 instances each.
While sigmoid uses the input scale 8 and output scale 14,  for tanh both scales are 8.
\item {\em Google-30}: It contains 99 sigmoid and 99 tanh layers, with 100 instances each.
While sigmoid uses the input scale 6 and output scale 14,  for tanh both scales are 6.
\item {\em Heads}: It contains 128 sigmoid and 128 tanh layers, with 18096 instances each.
While sigmoid uses the input scale 11 and output scale 14,  for tanh both scales are 11.
Additionally, the benchmark contains 8 sigmoid and 8 tanh layers, with 72384 instances each. 
For these layers, sigmoid uses the input scale 13 and output scale 14, and  for tanh both scales are 13.
Finally, it also contains 3 L2-Normalise layers
that have 1200, 1200 and 300 reciprocal square-root operations.
The layers have input scales 12, 10 and 12  and output scales 11, 9 and 11, respectively.    
\end{itemize}
\noindent Note that the Heads model makes about 3 million calls to sigmoid/tanh, which is three orders
of magnitude larger than the number of calls to these functions in the benchmarks used by prior work (\sectionref{priordnns}).

In~\tableref{rnns}, we present the latency and communication required
by \tool on above benchmarks.
Using our protocols, Industrial takes $4$ seconds, Google-30
takes under a minute, and Heads takes less than 7 minutes.
The time per inference can be further improved by {\em batching} multiple predictions.
For a batch size of $128$, the amortized time per inference of Industrial is
0.1s and of Google-30 is 1.1s! The savings in batching come from
amortizing the networking cost by
packing data from multiple inference queries.
 Owing to the high numerical precision of our math functionalities
 (\sectionref{validate}), \tool either matches or exceeds the model
 accuracy of the provided fixed-point ML model.
In Heads, about half the time is spent in math operations and the rest of the time is spent in matrix multiplications, convolutions, and Hadamard products.
The good performance on end-to-end benchmarks is a result of
co-designing precise math functionalities and efficient protocols.

Next, we perform an ablation study. In particular, the fixed-point code with our math functionalities can be run with other protocols. However, prior work on secure inference don't support 
juggling between different bitwidths that our math functionalities require. Hence, for running these functionalities with any prior protocol, we need to use an appropriately large uniform bitwidth.
We evaluate our benchmarks with ABY~\cite{aby} using the necessary  bitwidth of 64 as a baseline in~\tableref{rnns}.
 ABY~\cite{aby} provides general purpose state-of-the-art 2PC protocols that have been used by recent work on secure inference~\cite{hycc,ezpc,minionn,secureml}. We have added a new code generator to~\cite{shiftry} that generates {\sc{EzPC}}~\cite{ezpc} code  which is then automatically translated to ABY code.
Other generic protocols that have suitable frontends~\cite{oblivm,arm2gc,cbmcgc,fairplay}, like garbled circuits, are several orders of magnitude slower than ABY~\cite{secureml,ezpc}: ML inference involves many multiplications that are very expensive with garbled circuits.  
\tool is over $500\times$ better than ABY in communication and more than an order of magnitude faster in runtime. Without our protocols, it takes almost an hour to run Google-30. 
This situation is further exacerbated on bigger models and running the Heads model with ABY is intractable because it requires hundreds of terabytes of communication. 
With batching, the performance differences are stark: \tool is three orders of magnitude better in latency and communication compared to the ABY baseline.

\section{Other Related work}
\label{sec:related}
Prior 2PC works that use high degree
polynomials for approximating math functions~\cite{HemenwayLOW16,AlyS19,KimSKKHC19,raykovaspdz} need
degree 7 or higher to maintain accuracy.
In the course of this work, we have observed that evaluating polynomials with degree 3 or higher 
with 2PC is much more expensive than the LUT-based implementations
of~\sectionref{mathlib-protocols}.
Some prior works on secure inference implement math functions with ad
hoc approximations that can lose model accuracy:
e.g. SecureML~\cite{secureml} and ABY2.0~\cite{aby2} use a crude
3-piece linear approximation, Ball et al.~\cite{BallCMRS19} replace
tanh with the signum function, and Glyph~\cite{abs-1911-07101} and
Nandakumar et al.~\cite{NandakumarRPH19} use tables of approximate
results 
Most recent works on secure inference limit their evaluation to
benchmarks that don't use math
functions~\cite{delphi,cryptflow2,gazelle,chet,cryptonets,mpml,edabits}.
Prior 2PC works that use
     floating-point representations (instead of fixed-point representations) have much higher performance overheads\mbox{\cite{ChangL01,BaiYSLM18,FranzS11,emp-toolkit,abyfloat,LiuCHLW13,AliasgariB13,AliasgariBB17,mpspdz}}

Other relevant works that need additional parties to ensure security
such as 3PC with honest majority or 2PC with trusted dealer include
\cite{wenjie,Catrina18,CatrinaS10,Catrina19,Catrina19a,AliasgariBZS13,DimitrovKKRW16,AlyS19},
Chameleon~\cite{chameleon},
CrypTen~\cite{crypten},
TF-Encrypted~\cite{tfencrypted}, CrypTFlow~\cite{cryptflow},
PySyft~\cite{pysyft}, ABY$^3$~\cite{aby3},
SecureQ8~\cite{quantizednn}, and
Sharemind~\cite{sharemind-randmets,LaudR15,KammW15,PullonenS15,KripsW14}.
Some of these works have considered approximations to math functions
and, similar to 2PC works, they either use polynomial-based
approximations (e.g.~\cite{AlyS19,KripsW14,wenjie}) or work over
floating-point
(e.g.~\cite{Catrina19,Catrina19a,AliasgariBZS13,sharemind-randmets,LaudR15,KammW15,PullonenS15}).
Kerik et al. 
\cite{sharemind-randmets} also consider building blocks such as
extension, truncate-and-reduce, and multiplication of non-uniform
bitwidths in the 3PC context.
In terms of representations, while floating-point and fixed-point
representations are most common,~\cite{DimitrovKKRW16} proposed the
new representations of golden-section and logarithmic numbers and
evaluated using 3PC protocols.

Recent works on silent-OT~\cite{silentot,ferret} provide OT extensions with much lower communication than IKNP-style extensions~\cite{iknp}, at the cost of higher computational overhead. Since our protocols make use of OTs in a black-box manner, silent-OT can be used to obtain lower communication. However, in our setting, when the IKNP-OT instances are computed by multiple threads and are ``load-balanced" (i.e., each party plays the role of the sender in half the OT instances and as the receiver in the other half), we empirically observe that IKNP-style extensions are more performant than silent-OT in our LAN evaluation environment. Hence, \tool uses IKNP-style OT extensions in Section~\ref{sec:eval}.

\section{Conclusion}
\label{sec:concl}
We presented novel secure implementations of math functions that rely on cryptographic protocols for mixed-bitwidths.
These implementations, with up to $423\times$ lower communication than the state-of-the-art, 
help us evaluate ML models that have three orders of magnitude
more calls to math functions than benchmarks considered by prior work. 
Compared to a baseline, \tool
achieves three orders of magnitude lower communication and latency.
While prior work on secure 2-party inference has focused on image analysis,
\tool provides the first implementations of RNNs operating on speech data, sensor data,
and, in combinations with CNNs, detecting heads with state-of-the-art accuracy.
Because of high numerical precision of our math implementations, there is no loss in model accuracy over cleartext.
Although, in this work, we have focused on particular functions that occur in many ML models, the recipe of look ups followed by
Newton Raphson iterations to obtain precise functionalities is well-known in embedded systems and can be instantiated for other math functions as well.
We believe that our novel 2PC protocols would help provide the building blocks necessary for such functionalities.

\section*{Acknowledgement}\label{sec:ack}
\noindent
We thank Pratik Bhatu, Aayan Kumar, and Aditya Kusupathi for their help with the implementation and the evaluation.

\bibliographystyle{IEEEtranS}
\bibliography{main}

\appendix
\subsection{Optimized Protocol for $\fmux{\ell}$}
\label{app:mux}
In this section, we present an optimized protocol for $\fmux{\ell}$ which utilizes COT and builds over the protocol used in~\cite{cryptflow2}.
Our optimization relies on the following observation: consider $x \in \bbZ_2$ with shares $\share{x}{B}{} = (x_0, x_1)$ and $y \in \bbZ_L$ with shares $\share{y}{\ell}{} = (y_0, y_1)$, then the following holds:
\begin{align*}
    x \umult{\ell} y &= (x_0 \oplus x_1) \umult{\ell} (y_0 + y_1)\\
    &= (x_0 + x_1 - 2x_0 \umult{\ell} x_1) \umult{\ell} (y_0 + y_1) \\
    &= x_0 \umult{\ell} y_0 + x_1 \umult{\ell} (y_0 - 2x_0 \umult{\ell} y_0) \\
    & \quad + x_1 \umult{\ell} y_1 + x_0 \umult{\ell} (y_1 - 2x_1 \umult{\ell} y_1)
\end{align*}
In the above, the terms $x_0 \umult{\ell} y_0$ and $x_1 \umult{\ell} y_1$ can be locally computed by $P_0$ and $P_1$, respectively, while for the other two terms, we use $\iknpcot{\ell}$ protocol. In particular, to calculate shares of $x_1 \umult{\ell} (y_0 - 2x_0 \umult{\ell} y_0)$ term, $P_0$ acts as the sender with correlation $(y_0 - 2x_0 \umult{\ell} y_0)$ and $P_1$ acts as the receiver with choice bit $x_1$; similarly the term can be computed with  the sender and receiver roles reversed. Note that both the COTs can be done in parallel giving us a 2-round solution which communicates $2\ell$ less bits than prior approach from~\cite{cryptflow2} that used 2 instances of $\kkot{2}{\ell}$.

\subsection{Wrap and All Ones}
\label{app:wrap-eq}

Recall that the functionality $\fwrapeq{\ell}(x,y)$ outputs $(\share{w}{B}{} \concat \share{e}{B}{})$ such that $w = \wrap(x, y, L)$ and $e = \ind{(x + y \bmod L) = L-1}$. Consider the $\ell$-bit  functionality $\feq{\ell}(x,y)$  that returns $\share{e}{B}{}$ such that $e = \ind{x = y}$. 
Then, $\fwrapeq{\ell}(x,y) = \fmill{\ell}(L-1-x, y)\concat \feq{\ell}(L-1-x, y)$, that is, millionaires' and equality on the same inputs. 
Now, to construct an efficient protocol for $\fmill{\ell}$, CrypTFlow2~\cite{cryptflow2} used the following recurrence relations: Let $x = (x_1 \concat x_0)$ and $y = (y_1 \concat y_0)$ such that $x_i, y_i \in \zo^{\ell/2}$ for $i \in \zo$. Then,
\begin{align*}
    \ind{x < y} = \ind{x_1 &< y_1} \oplus (\ind{x_1 = y_1} \land \ind{x_0 < y_0}) \\
\ind{x = y} &= \ind{x_1 = y_1} \land \ind{x_0 = y_0}
\end{align*}

That is, they reduce the millionaires' on $\ell$-bit strings to millionaires' and equalities on smaller strings. While they computed millionaires' instances on all nodes, they skipped a small number of equality computations that were not used, \eg the root note. For $\fwrapeq{\ell}$, we compute millionaires' and equality on all notes and this marginally increases the cost over the protocol for $\fmill{\ell}$. Nonetheless, the communication cost of $\fwrapeq{\ell}$ is at most $\secpar\ell + 14\ell$.

\subsection{Truncation}
\label{app:truncation}

\subsubsection{Proof for \lemmaref{lrs}}

    For $b \in \zo$, let $x_b = \share{x}{\ell}{b}$. Over $\bbZ$, we can write $x_b = u_b \cdot 2^s + v_b$ and have the following:
\begin{align*}
x_0 + x_1 &= (v_0 + v_1) + 2^s (u_0 + u_1) \\
&= (v_0 + v_1 - c \cdot 2^s) + 2^s (u_0 + u_1 - d \cdot 2^{\ell-s}) \\
&\quad\quad + c \cdot 2^s + d \cdot L \\
&= v' + 2^s (u' + c) + d \cdot L
\end{align*}
    Let $w' = \ind{u' + c > 2^{\ell-s}-1}$. Then
\begin{align}
    x_0 + x_1 &= v' + 2^s (u' + c - w' \cdot 2^{\ell-s}) + L\cdot(d+w') \label{eq:lrs2-1}
\end{align}
    When $d = 1$, then $e = 0$ and $u' = u_0 + u_1 - 2^{\ell-s}$. Since $u_0, u_1 \leq 2^{\ell-s}-1$, we have that $u' \leq 2^{\ell-s}-2$. Therefore, $w' = 0$ (because $c \in \zo$). On the other hand when $d = 0$, $u' = u_0 + u_1 \leq 2^{\ell-s}-1$. Therefore, $w' = 1$ when $u' = 2^{\ell-s}-1$ (i.e., $e=1$) as well as $c = 1$, and 0 otherwise. Since at most one of $d$ and $w'$ is 1 in any given case, we can rewrite~\equationref{lrs2-1} as:
\begin{align*}
x_0 + x_1 &= v' + 2^s(u' + c - w' \cdot 2^{\ell-s}) + L\cdot(d \oplus (c \land e))
\end{align*}
Since $v' < 2^s$ and $u' + c - w' \cdot 2^{\ell-s} < 2^{\ell-s}$, $w = d \oplus (c \land e)$. \\

\subsubsection{Division by power-of-2}
We can write $\divtwo{x}{s} = (x \ggA s) + m_x \wedge c$, where $m_x = \ind{x \geq 2^{\ell-1}}$ is the MSB of $x$ and $c = \ind{x \bmod{2^{s}} \neq 0}$.
In this equation, $m_x$ can be computed with a call to $\fmill{\ell-1}$ using the integer DReLU protocol from \cite{cryptflow2} and $c$ can be computed with an equality check on $s$-bit inputs.
We get $m_x \wedge c$ in $\ell$-bits with a call each to $\fand{}$ and $\fBtoA{\ell}$, and then a final call to $\fars{\ell,s}$ gives us $\divtwo{x}{s}$.
Since we have already computed the MSB of $x$, we employ the MSB-to-wrap optimization (\sectionref{msb-to-wrap-opt}) here to minimize the cost of $\fars{\ell,s}$.
The exact cost expression for computing $\mathsf{DivPow2}$ is given in \tableref{comm-costs}.

\subsection{Multiplication}
\label{app:mult}
Here, we formally describe our protocols for cross term multiplication $\fcross{m,n}$ and matrix multiplication. \\

\subsubsection{Cross Term Multiplication, $\fcross{m, n}$}
\label{app:cross}
Our protocol for $\fcross{m, n}$ uses COT similar to prior works~\cite{secureml,aby,cryptflow2}, but unlike prior works, we support operands of different bitlengths. We present our protocol in~\algoref{cross} for the $m \leq n$ case. When $m > n$, we simply reverse the roles of the parties in our protocol so that only $n$ COTs are performed.
Correctness of this protocol follows similarly to the prior works. \\
\begin{algorithm}[t]
\caption{Cross Term Multiplication, $\protcross{\sint,\lint}$:}
\label{algo:cross}
\begin{algorithmic}[1]
\small

\Require $P_0$ holds $x \in \bbZ_M$ and $P_1$ holds ${y\in\bbZ_N}$, where ${m\leq n}$.
\Ensure $P_0$ \& $P_1$ get $\share{z}{\ell}{b}$, where $z = x \umult{\ell} y$ and $\ell = m + n$.

\vspace{0.1cm}
\State $\party{0}$ parses $x$ as an $\sint$-bit string $x~=~x_{\sint-1}\concat \cdots \concat x_{0}$, where $x_{i} \in \zo$.
\For{$i = \{0, \ldots, m-1\}$}
    \State $P_0$ \& $P_1$ invoke $\iknpcot{\ell-i}$, where $P_0$ is the sender with input $x_i$ and $P_1$ is the receiver with input $y$, and learn $\share{t_i}{\ell-i}{}$. \label{cross-terms-cot}
\EndFor
\State For $b \in \zo$, $P_b$ sets $\share{z}{\ell}{b} = \sum_{i=0}^{\sint-1} 2^{i} \cdot \share{t_i}{\ell-i}{b}$.
\end{algorithmic}
\end{algorithm}

\subsubsection{Matrix Multiplication}
\label{app:matrix-mult}
Before we look at matrix multiplication, we first set some notation starting with operator $\umatmul{\ell}: \bbZ^{d_1 \times d_2} \times \bbZ^{d_2 \times d_3} \rightarrow \bbZ^{d_1 \times d_3}_{L}$, which does a matrix multiplication between two input matrices $X$ and $Y$ such that $X \umatmul{\ell} Y = X \times Y \bmod{L}$ .
Similarly to the $\umult{\ell}$ notation, when one of the matrices has elements over ring $\bbZ_{M}$, we use the lossless typecast operator $\zeta_{m}$ to map all elements of that matrix to $\bbZ$.
All the \textit{single-input} functionalities we consider naturally extend to matrices, where the functionality is independently applied to all elements of the input matrix to output a matrix of the same dimensions.
The shares of a matrix $X \in \bbZ^{d_1 \times d_2}_{M}$ are denoted by $\share{X}{m}{}$, where $\share{X}{m}{} = \{ \share{X[i,j]}{m}{} \}_{i\in[d_1], j\in[d_2]}$, and the shares of its transpose are denoted by $\share{X^T}{m}{}$.
\par
Now, consider the matrix multiplication functionality $\fumatmul{m,n,d_1,d_2,d_3}$ that takes as input $\share{X}{m}{} \in \bbZ_{M}^{d_1 \times d_2}$ and $\share{Y}{n}{} \in \bbZ_{N}^{d_2 \times d_3}$ and outputs $\share{Z}{\ell}{} \in \bbZ_{\intx}^{d_1 \times d_3}$ such that $\ell = m + n + \lceil \log d_2 \rceil$ and $Z = X \umatmul{\ell} Y$.
As described in \sectionref{multiplication}, we need the additional $e = \lceil \log d_2 \rceil$ bits to prevent integer overflow due to additions.
When $m \leq n$, we extend the input matrix $\share{Y}{n}{}$ to get $\share{Y'}{n'}{}$ for $n' = n+e$.
Then, Equation~\ref{eq:umult} generalizes to matrices as follows: 

$X \umatmul{\ell} Y' = X_0 \umatmul{\ell} Y'_0 + X_1 \umatmul{\ell} Y'_1 + X_0 \umatmul{\ell} Y'_1 + X_1 \umatmul{\ell}~Y'_0 - 2^{n'} \umult{\ell} (X \umatmul{m} W_{Y'}) - M \umult{\ell} (W_{X} \umatmul{n'} Y')$, 
where ${W_X~=\wrap(X_0, X_1, M)}$ and $W_{Y'} = \wrap(Y'_0, Y'_1, 2^{n'})$.

Similar to $\fcross{m,n'}$, we define a functionality $\fmatcross{m,n',d_1,d_2,d_3}$ for matrices to compute the cross-terms $X_0 \umatmul{\ell} Y'_1$ and $X_1 \umatmul{\ell} Y'_0$.
This functionality can be realized naively by making $d_1 d_2 d_3$ independent calls to $\protcross{m,n'}$.
Instead, we can do much better by observing that in a matrix multiplication, each element of $X$ is multiplied with $d_3$ elements of $Y$.
Thus, rather than doing $d_3$ independent COTs on $\ell-i$ bit-strings in Step~\ref{cross-terms-cot} of $\protcross{m,n'}$, we can perform a single COT on $d_3 \cdot (\ell-i)$ bit-strings (while respecting the independent correlations).
This method of batching COTs was also used in prior works on secure inference~\cite{secureml,cryptflow2}, and it leads to an overall communication of $d_1 d_2 (m \lambda + (mn' + m^2/2 + m/2)d_3)$ bits.

Note that $\share{W_{X}}{B}{}$ and $\share{W_{Y'}}{B}{}$ can be computed by making $d_1 d_2$ calls to $\fwrap{m}$ and $d_2 d_3$ calls to $\fwrap{n'}$, respectively.
Since the terms $X_i \umatmul{\ell} Y'_i$ can be computed locally, the only terms left to compute are $X \umatmul{m} W_{Y'}$ and $W_{X} \umatmul{n'} Y'$.
They can be computed using the following functionality $\fblah{\ell, d_1, d_2, d_3}$ that takes a bit-matrix $\share{W}{B}{} \in \zo^{d_1 \times d_2}$ and a matrix $\share{X}{\ell}{} \in \bbZ_{L}^{d_2 \times d_3}$ as inputs, and outputs a matrix $\share{Z}{\ell}{} \in \bbZ_{L}^{d_1 \times d_3}$ such that $Z = W \umatmul{\ell} X$.
We use the OT-based MUX protocol from \cite{cryptflow2} to implement $\fblah{\ell, d_1, d_2, d_3}$, and also leverage the batching technique here to reduce the number of OTs.
The communication required by this protocol is $2 d_1 d_2 (\secpar + 2 \ell d_3)$~bits.
\par
Our complete protocol for $\fumatmul{m,n,d_1,d_2,d_3}$ is presented in \algoref{umatmul} for the $m \leq n$ case.
The total communication cost of this protocol is ${d_1d_2d_3 ((2m+4)(n+e) + m^2 + 5m)} + {d_1d_2 (\secpar(3m+6)+14m + e - 6)} + {d_2d_3(\secpar (n + 2) + 14n)}$ bits.
In the protocol, we extend $Y$ because it has elements of larger bitwidth, and this strategy leads to better overall communication in most cases.
The other case of $m > n$ is similar and we extend the entries of matrix $X$ by $e$ bits. 

\begin{algorithm}[t]
\caption{Unsigned Matrix Multiplication, $\protumatmul{\sint,\lint,d_1,d_2,d_3}$:}
\label{algo:umatmul}
\begin{algorithmic}[1]
\small
\Require $P_0$ \& $P_1$ hold $\share{X}{m}{}$ and $\share{Y}{n}{}$, where ${X \in \bbZ_{M}^{d_1 \times d_2}}$, ${Y \in \bbZ_{N}^{d_2 \times d_3}}$ and $m \leq n$.
\Ensure $P_0$ \& $P_1$ get $\share{Z}{\ell}{}$, where $Z = X \umatmul{\ell} Y$, ${\ell = m + n + e}$ and $e = \lceil \log d_2 \rceil$.

\vspace{0.1cm}
\State $P_0$ \& $P_1$ invoke $\fzxt{n, n+e}(\share{Y}{n}{})$ and learn $\share{Y'}{n'}{}$.
\State For $b \in \zo$, let $X_b = \share{X}{\sint}{b}$ and $Y'_b = \share{Y'}{n'}{b}$.
\State $P_0$ and $P_1$ invoke the following functionalities.
\Indent
\State $\fmatcross{m, n', d_1, d_2, d_3}(X_0, Y'_{1})$ and learn $\share{C}{\ell}{}$.
\vspace{0.05cm}
\State $\fmatcross{n', m, d_3, d_2, d_1}({Y'}_{0}^{T}, X_1^{T})$ and learn $\share{D}{\ell}{}$.
\vspace{0.05cm}
\State $\fwrap{m}(X_0, X_1)$ to learn $\share{W_X}{B}{}$.
\vspace{0.05cm}
\State $\fwrap{n'}(Y'_0, Y'_1)$ to learn $\share{W_{Y'}}{B}{}$.
\vspace{0.05cm}
\State $\fblah{m, d_3, d_2, d_1}(\share{W^T_{Y'}}{B}{}, \share{X^T}{m}{})$ to learn $\share{G}{\sint}{}$.
\vspace{0.05cm}
\State $\fblah{n', d_1, d_2, d_3}(\share{W_X}{B}{}, \share{Y'}{n'}{})$ to learn $\share{H}{n'}{}$.
\EndIndent
    \State $P_b$ outputs  $X_b \umatmul{\ell} Y'_b + \share{C}{\ell}{b} + \share{D^T}{\ell}{b} - 2^{n'} \umult{\ell} \share{G^{T}}{\sint}{b} - {2^{m} \umult{\ell} \share{H}{n'}{b}}$ for $b \in \zo$.

\end{algorithmic}
\end{algorithm}

\begin{table*}[t]
\centering
\renewcommand*{\arraystretch}{1.5}
\begin{tabular}{|c|c|c|}
\hline
    Protocol  & Comm. (bits)               & Rounds \\ \hline \hline
    $\protzxt{m,n}$ \& $\protsxt{m,n}$ & $\secpar (m + 1) + 13 m + n$ & $\log m + 2$ \\ \hline
    $^\star\protzxt{m,n}$ \& $^\star\protsxt{m,n}$ & $2\secpar - m + n + 2$ & $4$ \\ \hline
    $\protlrs{\ell,s}$ \& $\protars{\ell,s}$ & $\secpar(\ell + 3) + 15\ell + s + 20$ & $\log\ell + 3$ \\ \hline
    $^\star\protlrs{\ell,s}$ \& $^\star\protars{\ell,s}$ & $\secpar(s + 3) + \ell + 15s + 2$ & $\log s + 2$ \\ \hline
    $\protr{\ell,s}$ & $\secpar (s + 1) + \ell + 13 s$ & $\log s + 2$ \\ \hline
    $\protdivtwo{\ell,s}$ & $\secpar(\ell + 7s/4 + 4) + 16\ell + 23s - 5$ & $\log\ell + 4$ \\ \hline
    $\protumult{m,n}$ \& $\protsmult{m,n}$ & $\secpar(3\mu + \nu + 4) + 2\mu\nu + \mu^2 + 17\mu + 16\nu$ & $\log\nu + 2$ \\ \hline
    $^\star\protumult{m,n}$ \& $^\star\protsmult{m,n}$ & $\secpar (2\mu + 6) + 2\mu\nu + \mu^2 + 3\mu + 2\nu + 4$ & $4$ \\ \hline
    $\protdigdec{\ell,d}$ & $(\ell/d - 1) (\secpar (d + 2) + 15 d + 20)$ & $\log d + \ell/d + 1$ \\ \hline
    $\protmsnzb{\ell,d}$ & $(\ell/d - 1) (\secpar (d + 8) + 2^d (\iota + 1) + 15d + 2 \iota + 60) + 6 \secpar + 2^d (\iota + 1) + \ell^2 + 2 \iota$ & $\log d + 2\ell/d + 7$ \\ \hline
\end{tabular}
    \caption{Exact communication and round expressions for our building blocks, assuming that the cost of $\protmill{\ell}$ and $\protmilleq{\ell}$ is $\secpar \ell + 14 \ell$ bits. $\mu = \mathsf{min}(m,n), \nu = \mathsf{max}(m,n)$, and $\star$ denotes the variant of the protocol in which the MSBs of the inputs are already known in the clear. In case the MSBs are known in the shared form, the additional cost is just $\secpar + 2$ bits per input.}
\label{tab:comm-costs}
\end{table*}

\subsection{Digit Decomposition} 
\label{app:digit-decomp}

We consider the functionality $\fdigdec{\ell, \{d_i\}_{i \in [c]}}$ that decomposes an $\ell$-bit number into $c$ sub-strings or digits of lengths $\{d_i\}$ such that $\sum_{i \in [c]} d_i = \ell$. 
More formally, $\fdigdec{\ell,  \{d_i\}_{i \in [c]}}$ takes $\share{x}{\ell}{}$ as input and outputs 
$\share{z_{c-1}}{{d_{c-1}}}{}, \ldots, \share{z_0}{{d_0}}{}$ 
such that $x = z_{c-1} \concat \ldots \concat z_0$.
We use this functionality in extracting digits to be used as input to lookup tables for approximations for exponential, initial approximation of reciprocal in sigmoid/tanh and reciprocal square root. 

For ease of exposition we first consider a simplified functionality $\fdigdec{\ell,d}$ with $d \mid \ell$ that outputs $c = \ell/d$ digits of equal length $d$ and present our protocol for this functionality in ~\algoref{digdec}. Idea is as follows: To compute the shares of $z_i$, it suffices to compute the carry of lower bits into this digit when reconstructing shares of $x$. That is, consider a parsing of $\ell$-bit string $\share{x}{\ell}{b}$ as $y_{b,c-1}\concat \ldots \concat y_{b,0}$ such that $y_{b,i} \in \zo^\digit$ for all $i \in [c]$ for $b \in \zo$. Also, set $Y_{b,i} =  y_{b,i}\concat \ldots \concat y_{b,0}$ for all $i \in [c]$, $b \in \zo$. Now, observe that $z_i = y_{0,i} + y_{1,i} + \carry_{i} \bmod 2^d$, where $\carry_{i} = Y_{0, i-1} + Y_{1, i-1} \geq 2^{id}$. Alternatively, $\carry_{i} = \wrap(Y_{0,i-1}, Y_{1,i-1}, 2^{id})$. In our protocol, we compute this $\carry_i$ using \lemmaref{lrs} iteratively (similar to our protocol for $\flrs{\ell,s}$) and the variable $u_i$ corresponds to $\carry_i$.
The communication complexity of our protocol for the simplified setting is $(c - 1) (\secpar (d + 2) + 15 d + 20)$ bits.

Also, it is easy to see that the above protocol generalizes to the case of unequal size digits, by parsing the initial shares appropriately and doing the same computation.
The communication for the generalized case is ${\sum_{i \in [c-1]} (\secpar (d_i + 2) + 15 d_i + 20)}$ bits. %
In contrast, doing a digit-decomposition using GC would require $\secpar (6\ell - 2c - 2)$ bits of communication. For example, for $\ell = 32$ and $d = 8$, our protocol has an improvement of $5.5\times$ over GC.

\begin{algorithm}[t]
\caption{Digit Decomposition, $\protdigdec{\ell,\digit}$:}
\label{algo:digdec}
\begin{algorithmic}[1]
\small

\Require $P_0$ \& $P_1$ hold $\share{x}{\ell}{}$ s.t. $c = \ell/\digit$.
\Ensure $P_0$ \& $P_1$ get $\{\share{z_i}{\digit}{}\}_{i \in [c]}$ s.t. $x = z_{c-1} \concat \ldots \concat z_0$.

\vspace{0.1cm}
\State For $b \in \zo$, $P_b$ parses $\share{x}{\ell}{b}$ as an $\ell$-bit string $y_{b, c-1} \concat \ldots \concat y_{b,0}$ s.t. $y_{b,i} \in \zo^\digit$ for all $i \in [c]$.
    \State For all  $i \in \{0,\ldots,c-2\}$,   \alice \& \bob invoke $\fwrapeq{\digit}(y_{b,i}, y_{b,1})$ and learn $\share{w_i}{B}{}\concat \share{e_i}{B}{}$.
\State For $b \in \zo$, $P_b$ sets $\share{u_0}{B}{b} = 0$ and $\share{z_0}{\digit}{b} = y_{b,0}$.
\For{$i \in \{1,\ldots,c-1\}$}
    \State \alice \& \bob invoke $\fand(\share{u_{i-1}}{B}{},\share{e_{i-1}}{B}{})$ to learn $\share{v_{i-1}}{B}{}$.
    \State For $b \in \zo$, $\party{b}$ sets $\share{u_i}{B}{b} = \share{v_{i-1}}{B}{b} \oplus \share{w_{i-1}}{B}{b}$.
    \State \alice \& \bob invoke $\fBtoA{\digit}(\share{u_i}{B}{})$ and learn $\share{u_i}{\digit}{}$.
    \State For $b \in \zo$, $\party{b}$ sets $\share{z_i}{\digit}{b} = y_{b,i} + \share{u_i}{B}{b}$.
\EndFor

\end{algorithmic}
\end{algorithm}

\subsection{Most Significant Non-zero Bit (MSNZB)} 
\label{app:msnzb}
For an $\ell$-bit integer $x$, $\msnzb(x)$ refers to the index of the most significant non-zero-bit. That is, $\msnzb(x) = k \in [\ell]$, if $x_k = 1$ and $x_j = 0$ for all $j > k$. Alternatively, $\msnzb(x) = k$ if and only if $2^{k} \leq x < 2^{k+1}$. For the special case of input being 0, $\msnzb(0) = 0$.
Consider the functionality $\fmsnzb{\ell}$ that takes as input $\share{x}{\ell}{}$ and outputs $\{\share{z_i}{B}{}\}_{i \in [\ell]}$ such that $z_i = 1$ if $\msnzb(x) = i$ and $0$ otherwise.
Our protocol for $\fmsnzb{\ell}$ reduces to MSNZB-like computation on integers on smaller bit-length as follows: For simplicity of exposition, consider $d \in \mathbb{N}$ such that $d \mid \ell$. First, we invoke $\fdigdec{\ell,d}$ to decompose $\ell$-bit integer $x$ into $c = \ell/d$ integers of $d$-bits, say $\{y_i\}_{i \in [c]}$. Now, we compute MSNZB on each of these smaller integers $y_i$ by taking into account their position $i$ in $x$ and output an index in $[\ell]$ which corresponds to $\msnzb(y_i) + i \cdot d$. Note that $\msnzb(x) = \msnzb(y_i) + i \cdot d$ if $y_i \ne 0$ and $y_j = 0$ for all $j > i$. To realize this logic we also compute whether $y_i = 0$ for all $i \in [c]$. 

More formally, let $\iota = \log \ell$ and consider the functionality $\fmsnzbproj{d, \ell, i}$ for $i \in [c]$ that takes as input $\share{y}{d}{}$ and outputs $\share{u}{\iota}{}$ such that $2^{u-id} \leq y < 2^{u-id+1}$. Also, consider $\fzeros{d}$ functionality that takes as input $\share{y}{d}{}$ and outputs $\share{v}{B}{}$ such that $v = \ind{y = 0}$. First, our protocol invokes $\fmsnzbproj{d, \ell, i}$ on each of $\share{y_i}{d}{}$ (obtained from $\fdigdec{\ell,d}(\share{x}{\ell}{})$) to learn $\share{u_i}{\iota}{}$. Next, we invoke $\fzeros{d}(y_i)$ to learn $\share{v_i}{B}{}$. 
Now, for all $i \in [c]$, we compute  $z_i' = u_i \cdot (1 \oplus v_i) \cdot \prod_{j > i} v_j$. 
Note that $z'_i = u_i$ if  $y_i \ne 0$ and $y_j = 0$ for all $j > i$ and $0$ otherwise. Moreover, at most one $z'_i$ is non-zero. Hence, we compute $\msnzb(x) = \tilde{z} = \sum_i z'_i$. 
Finally, to output the one-hot encoding described above, we invoke the functionality $\fonehot{\ell}$ that takes as input $\share{\tilde{z}}{\iota}{}$ and outputs $\{\share{z_i}{B}{}\}_{i \in [\ell]}$ such that $z_{i} = 1$ for $i = \tilde{z}$ and 0 otherwise.
We present our protocol for $\fmsnzb{\ell}$ in~\algoref{msnzb}, for the special case of $\digit \mid \ell$; it is easy to see that the general case works in a similar manner.
Our protocol makes $1$ call to $\fdigdec{\ell,d}$, $c$ calls each to $\fmsnzbproj{d, \ell, i}$, $\fzeros{d}$ (with $i$ going from 0 to $c-1$) and $\fmux{\iota}$, $2c-2$ calls to $\fand$ and $1$ call to $\fonehot{\ell}$.

We implement both $\fmsnzbproj{d, \ell, i}$ and $\fzeros{d}$ using LUTs with $d$-bit inputs. Moreover, since these are invoked on same input, we combine them into a single LUT with entries $(u_i\concat v_i)$.
Finally, we implement $\fonehot{\ell}$ using an LUT with $\iota$-bit input and $\ell$-bit entries.
The exact expression for communication for $d \mid \ell$ is given in \tableref{comm-costs}.
The expression for the general case can be computed similarly using expression in digit decomposition.
Based on empirical findings, we use $d = 8$ in our implementation.

\begin{algorithm}[t]
\caption{Most Significant Non-Zero Bit, $\protmsnzb{\ell,\digit}$:}
\label{algo:msnzb}
\begin{algorithmic}[1]
\small

\Require For $b \in \zo$, $\party{b}$ holds $\share{x}{\ell}{b}$, $c = \ell/d, \iota = \log \ell$.
\Ensure For $b \in \zo$, $\party{b}$ learns $\{\share{z_i}{B}{b}\}_{i \in [\ell]}$ s.t. $z_i = 1$ if $2^{i} \leq x < 2^{i+1}$ and 0 otherwise.
\vspace{0.1cm}

    \State $P_0$ $\&$ $P_1$ invoke $\fdigdec{\ell, \digit}(\share{x}{\ell}{})$ and learn $\{\share{y_i}{\digit}{}\}_{i \in [c]}$. 
    \For {$i \in \{0,\ldots,c-1\}$}
    \State $P_0$ $\&$ $P_1$ invoke $\fmsnzbproj{d,\ell,i}(\share{y_i}{\digit}{})$ and learn $\share{u_i}{\iota}{}$. \label{step:msnzb-proj}
    \State $P_0$ $\&$ $P_1$ invoke $\fzeros{\digit}(\share{y_i}{\digit}{})$ and learn $\share{v_i}{B}{}$. \label{step:msnzb-zeros}
    \State For $b \in \zo$, $P_b$ sets $\share{v'_{i}}{B}{b} = (b \oplus \share{v_{i}}{B}{b})$.
    \EndFor
    \State $P_0$ $\&$ $P_1$ invoke $\fmux{\iota}(\share{v_{c-1}'}{B}{}, \share{u_{c-1}}{\iota}{})$ and learn $\share{z_{c-1}'}{\iota}{}$.
    \State For $b \in \zo$, $P_b$ sets $\share{w_{c-1}}{B}{b} = b$.
    \For {$i \in \{c-2, \ldots, 0\}$}
    \State $P_0$ $\&$ $P_1$ invoke $\fand(\share{w_{i+1}}{B}{}, \share{v_{i+1}}{B}{})$ and learn $\share{w_i}{B}{}$.
    \State $P_0$ $\&$ $P_1$ invoke $\fand(\share{w_i}{B}{}, \share{v_i'}{B}{})$ and learn $\share{w_i'}{B}{}$.
    \State $P_0$ $\&$ $P_1$ invoke $\fmux{\iota}(\share{w_i'}{B}{}, \share{u_i}{\iota}{})$ and learn $\share{z_i'}{\iota}{}$.
    \EndFor
    \State For $b \in \zo$, $P_b$ sets $\share{\tilde{z}}{\iota}{b} = \sum_{i=0}^{c-1} \share{z_i'}{\iota}{b}$.
    \State $P_0$ $\&$ $P_1$ invoke $\fonehot{\ell}(\share{\tilde{z}}{\iota}{})$ and learn $\{\share{z_i}{B}{}\}_{i \in [\ell]}$. \label{step:msnzb-onehot}

\end{algorithmic}
\end{algorithm}

\end{document}